\documentclass[acmsmall]{acmart}

\setcopyright{none}
\renewcommand\footnotetextcopyrightpermission[1]{} %
\usepackage{subfigure}
\usepackage{balance}
\usepackage{graphicx}
\usepackage{multirow}
\usepackage{amssymb}
\usepackage{wrapfig}
\usepackage{alltt}
\usepackage{xspace}
\usepackage{verbatim}
\usepackage{epstopdf}
\usepackage{rotating}
\usepackage{subfigure}
\usepackage{rotating,booktabs,makecell}
\usepackage{slashbox}
\usepackage{mathtools}%
\usepackage{listings}
\usepackage{tablefootnote,footnote}
\usepackage{tikz}
\usepackage[keeplastbox]{flushend}
\usetikzlibrary{shapes,snakes}
\usepackage{color, colortbl}

\makeatletter
\let\@authorsaddresses\@empty
\makeatother

\newcommand{\descr}[1]{\smallskip\noindent\textbf{\small #1}}

\usepackage[hang,flushmargin,stable]{footmisc}
\renewcommand{\footnotesize}{\fontsize{8}{9}\selectfont}
\usepackage{indentfirst}

\makeatletter
\def\url@foostyle{%
  \@ifundefined{selectfont}{\def\UrlFont{\rm}}{\def\UrlFont{\rmfamily}}}
\makeatother
\newif\ifcomment
\commenttrue 
\ifcomment
	\newcommand{\edc}[1]{\textbf{\em\color{red}#1}}
	\newcommand{\lucky}[1]{\textbf{\em\color{blue}#1}}
	\newcommand{\gs}[1]{\textbf{\em\color{purple}#1}}
	\newcommand{\enm}[1]{\textbf{\em\color{brown}#1}}
\else  
    \newcommand\edc[1]{}
    \newcommand\lucky[1]{}
    \newcommand\gs[1]{}
    \newcommand\enm[1]{}
\fi

\tikzstyle{highlighter} = [
  yellow,
  line width = \baselineskip,
]
\newcounter{highlight}[page]

\AtBeginShipout{\AtBeginShipoutUpperLeft{\ifthenelse{\value{highlight} > 0}{\tikz[remember picture, overlay]{\foreach \stroke in {1,...,\arabic{highlight}} \draw[highlighter] (begin highlight \stroke) -- (end highlight \stroke);}}{}}}


\makeatletter
\newcommand\smallscriptsize{\@setfontsize\scriptsize{5.75}{6.75}}
\makeatother

\newcommand{\approach}{\textsc{MaMaDroid}\xspace}
\newcommand{\droid}{\textsc{DroidAPIMiner}\xspace}
\newcommand{\FAM}{\textsc{FAM}\xspace}

\begin{document}

\title[MaMaDroid (Extended Version)]{MaMaDroid: Detecting Android Malware by Building Markov Chains of Behavioral Models (Extended Version)}
\thanks{This is an extended work of the authors' prior publication presented at NDSS 2017~\cite{mariconti2016mamadroid}. Work done while Gianluca Stringhini was at UCL. Submitted: November 2017, Accepted: February 2019.}

\author{Lucky Onwuzurike}
\author{Enrico Mariconti}
 \affiliation{
 \institution{University College London}}
\author{Panagiotis Andriotis}
 \affiliation{
 \institution{University of the West of England}}
\author{Emiliano De Cristofaro}
\author{Gordon Ross}
 \affiliation{
 \institution{University College London}}
\author{Gianluca Stringhini}
 \affiliation{
 \institution{Boston University}}

\begin{abstract}
As Android has become increasingly popular, so has malware targeting it, thus pushing the research community to propose different detection techniques. However, the constant evolution of the Android ecosystem, and of malware itself, makes it hard to design robust tools that can operate for long periods of time without the need for modifications or costly re-training. Aiming to address this issue, we set to detect malware from a behavioral point of view, modeled as the sequence of abstracted API calls. We introduce MaMaDroid, a static-analysis based system that abstracts the API calls performed by an app to their class, package, or family, and builds a model from their sequences obtained from the call graph of an app as Markov chains. This ensures that the model is more resilient to API changes and the features set is of manageable size. We evaluate MaMaDroid using a dataset of 8.5K benign and 35.5K malicious apps collected over a period of six years, showing that it effectively detects malware (with up to 0.99 F-measure) and keeps its detection capabilities for long periods of time (up to 0.87 F-measure two years after training). We also show that MaMaDroid remarkably outperforms DroidAPIMiner, a state-of-the-art detection system that relies on the frequency of (raw) API calls. Aiming to assess whether MaMaDroid?s effectiveness mainly stems from the API abstraction or from the sequencing modeling, we also evaluate a variant of it that uses frequency (instead of sequences), of abstracted API calls. We find that it is not as accurate, failing to capture maliciousness when trained on malware samples that include API calls that are equally or more frequently used by benign apps.

\end{abstract}

\maketitle

\thispagestyle{empty}

\section{Introduction}
\label{sec:introduction}
Malware running on mobile devices can be particularly lucrative, as it can enable attackers to defeat two-factor authentication for financial and banking systems~\cite{android2fa} %
and/or trigger the leakage of sensitive information~\cite{gordon2015information}.
As a consequence, the number of malware samples has skyrocketed in recent years and, due to its increased popularity, cybercriminals have increasingly targeted the Android ecosystem~\cite{androidtrend}.
Detecting malware on mobile devices presents additional challenges compared to desktop/laptop computers; smartphones have limited battery life, making it impossible to use traditional approaches
requiring constant scanning and complex computation~\cite{polakis2015powerslave}.
Thus, Android malware detection is typically performed in a centralized fashion, 
i.e., by analyzing apps submitted to the Play Store using 
Bouncer~\cite{oberheide2012dissecting}. However, many malicious apps manage to avoid detection~\cite{trusted,fake-wa},
and manufacturers as well as users can install apps that
come from third parties, whom may not perform any malware checks at all~\cite{Zhou2012hey}. 

As a result, the research community has proposed a number of techniques to detect malware on Android. %
Previous work has often relied on the permissions requested by apps~\cite{Enck2009,Sarma2012},
using models built from malware samples. This, however, is prone to false positives, since there are often legitimate reasons for 
benign apps to request permissions classified as dangerous~\cite{Enck2009}.
Another approach, used by \droid~\cite{Aafer2013DroidAPIMiner}, is to perform  classification based on API calls frequently used by malware. However, relying on the most common calls observed during training prompts the need for constant retraining, due to the evolution of malware and the Android API alike. For instance, ``old'' calls are often deprecated with new API releases, so malware developers may switch to different calls to perform similar actions.

\descr{MaMaDroid.} In this paper, we present a novel malware detection system for Android
that relies on the {\em sequence} of {\em abstracted} API calls performed by an app 
rather than their use or frequency, aiming to capture the behavioral model of the app.
We design \approach %
to abstract API calls to either the {\em class} name (e.g., {\tt java.lang.Throwable}) of the call or its {\em package} name %
(e.g., {\tt java.lang}) or its source (e.g.,
{\tt java}, {\tt android}, {\tt google}),  which we refer to as {\em family}. %

Abstraction provides resilience to API changes in the Android framework as families and packages are added and removed 
less frequently than single API calls.
At the same time, this does not abstract away the behavior of an app. 
For instance, packages include classes and interfaces used to perform similar operations
on similar objects, so we can model the types of operations from the package name alone. 
For example, %
the {\tt java.io} package is used for system I/O and access to the file system, even though there are different classes and interfaces provided by the package for such operations. 

After abstracting the calls, \approach %
analyzes the {\em sequence} of API calls performed by the app aiming to model the app's behavior %
using Markov chains. %
Our intuition is that malware may use calls for different operations, %
and in an order different
from benign apps. For example, android.media.MediaRecorder can be used by any app that has permission to record audio, 
but the call sequence may reveal that malware only uses calls from this class {\em after} calls to getRunningTasks(), 
which allows recording conversations~\cite{zhang2015leave}, as opposed to benign apps where calls from the class may appear in {\em any} order. 
Relying on the sequence of abstracted calls allows us to model behavior in a more complex way
than previous work, which only looked at the presence or absence of certain API calls
or permissions~\cite{Aafer2013DroidAPIMiner,arp2014drebin}, while still keeping the problem
tractable~\cite{ComplMacLear}.
\approach then builds a statistical model to represent the transitions between the API calls performed by an app
as Markov chains, and uses them to extract features. Finally, it classifies an app as either malicious or benign using the features it extracts from the app. 

\descr{Evaluation.} We present a detailed evaluation of the classification accuracy (using F-measure, Precision, and Recall) and runtime performance of \approach using a
dataset of almost 44K apps (8.5K benign and 35.5K malware samples). 
We include a mix of older and newer apps, from October 2010 to May 2016, 
verifying that our model is robust to changes in Android malware samples and APIs.
Our experimental analysis shows that \approach can effectively model both benign and malicious Android apps, and %
efficiently classify them. Compared to other systems such as \droid~\cite{Aafer2013DroidAPIMiner}, 
our approach allows us to account for changes in the Android
API, without the need to frequently retrain the classifier. 
Also, to the best of our knowledge,
our evaluation is done on one of the largest Android malware datasets used in a research paper.

To assess the impact of abstraction and Markov chain modeling on \approach, we not only compare to \droid, but also build a variant (called \FAM) that still abstracts API calls but instead of building a model from the sequence of calls, it does so on the frequency of calls, similar to \droid. 

Overall, we find that \approach can effectively detect unknown malware samples not only in the ``present,'' (with F-measure up to 0.99) but also consistently over the years
(i.e., when the system is trained on older samples and evaluated over newer ones), 
as it keeps an average detection accuracy, evaluated in terms of F-measure, of 0.87 after one year and 0.75 after two years (as opposed to
0.46 and 0.42 achieved by \droid~\cite{Aafer2013DroidAPIMiner} and 0.81 and 0.76 by \FAM). We also highlight that when the system is not efficient anymore (when the test set is newer than the training set by more than two years), it is as a result of \approach having low Recall, but maintaining high Precision.
We also do the opposite, i.e., training on newer samples and verifying that the system can still detect old malware.
This is particularly important as it shows that \approach can detect newer threats, while still identifying malware samples that have been in the wild for some time.

\descr{Summary of Contributions.} This paper makes several contributions. First, we introduce a novel malware detection approach implemented in a tool called \approach,  %
by abstracting API calls to their class, package, and family, and %
model the behavior of the apps through the sequences of API calls as Markov chains. %
Second, we can detect unknown samples from the same year as those used in training with an F-measure of 0.99, and also years after training the system, meaning that \approach does not need continuous re-training. Compared to previous work, %
\approach achieves higher accuracy with reasonably fast running times, while also being more robust to evolution in malware development and changes in the Android API. Third, %
by abstracting API calls and %
using frequency analysis %
we still perform better than a system that also uses frequency analysis but without abstraction (\droid). 
Fourth, we explore the detection performance of a finer-grained abstraction and %
show that abstracting to classes does not perform better than abstracting to packages. Finally, we make the code of \approach as well as the hash of the samples in our datasets publicly available\footnote{\url{https://bitbucket.org/gianluca_students/mamadroid_code}} and, on request, the apk samples, parsed call graphs, and abstracted sequences of API calls on which \approach has been evaluated on.

Note that this is an extended work of {\bf\em our prior publication presented at NDSS 2017}~\cite{mariconti2016mamadroid};
compared to that paper, here we make two main additional contributions:
(1) We present and evaluate a finer-grained level of API call abstraction---to {\em classes}, rather than packages or families;
(2) We introduce and assess a modeling approach based on the {\em frequency} rather than the sequences of abstracted API calls. We compare this to that presented in~\cite{mariconti2016mamadroid} as well as to \droid~\cite{Aafer2013DroidAPIMiner}, as the latter {\em also} uses the frequency of non-abstracted API calls.

\descr{Paper Organization.} %
Next section presents \approach,
then, Section~\ref{sec:data} introduces the datasets used throughout the paper. In Section~\ref{sec:markoveval}, we evaluate \approach in family and package modes, while in Section~\ref{sec:classeval}, we explore the %
effectiveness of finer-grained abstraction (i.e., class mode).
In Section~\ref{sec:evaluation}, we present and evaluate the variant using a frequency analysis model (\FAM), while we analyze runtime performances in Section~\ref{sec:runtime}. Section~\ref{sec:discussion} 
further discusses our results as well as its limitations. After reviewing related work in Section~\ref{sec:related}, the paper concludes in Section~\ref{sec:conclusion}.

\section{The MaMaDroid System}
\label{sec:method}
In this section, we introduce \approach, an Android malware detection system  
that relies on the transitions between different API calls performed by Android apps.

\subsection{Overview}
\approach builds a model of the sequence of API calls as Markov chains, which are in turn used to extract features for machine learning algorithms to classify apps as benign or malicious. 

\begin{figure}[h]
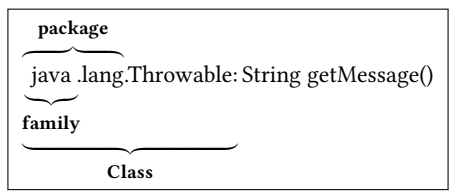

\small
$$\boxed{~~~\underbrace{\overbrace{\underbrace{\mbox{java}}_\text{\bf family}\mbox{\hspace{-0.1cm}.lang}}^\text{\bf package}\mbox{\hspace{-0.05cm}.Throwable:}}_\text{\bf Class}\mbox{String getMessage()}~~~}$$
\vspace{-0.25cm}
\caption{An example of an API call and its family, package, and class.}\label{fig:api}
\vspace{-0.15cm}
\end{figure}

\descr{Abstraction.} \approach does not actually use the {\em raw}
API calls, but abstracts each call to its family, package, or class.
For instance, the API call getMessage() in Fig.~\ref{fig:api} is parsed to, respectively, {\tt java}, {\tt java.lang}, and {\tt java.lang.Throwable}.\vspace{-0.15cm}

Given the three different types of abstractions, \approach operates in one of three modes, each using one of the types of abstraction. 
Naturally, we expect that the higher the abstraction, the lighter the system is, 
although possibly less accurate.

\begin{figure}[t]
 \center
 \includegraphics[width=0.9\textwidth]{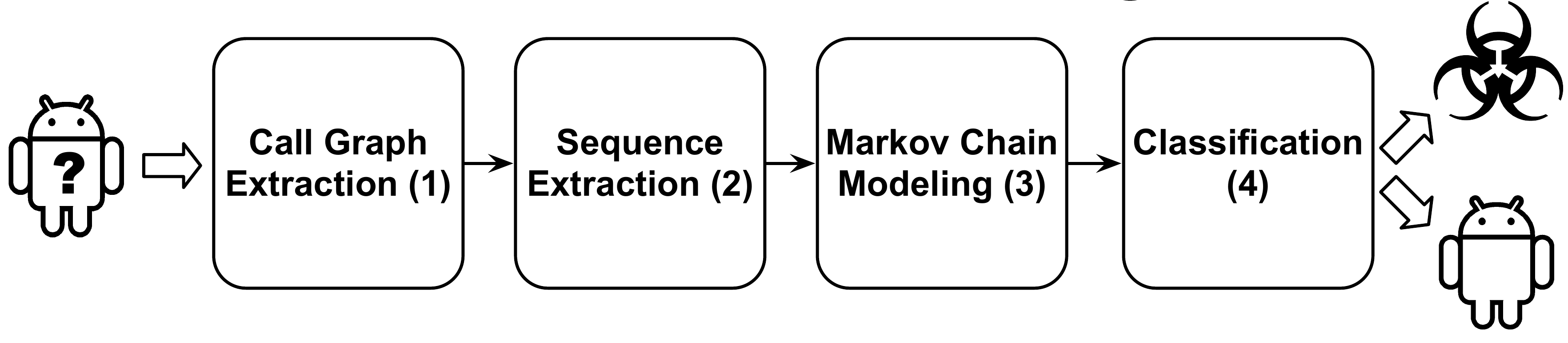}
 \vspace{-0.5cm}
 \caption{Overview of \approach operation. In (1), it extracts the call graph from an Android app, next, it builds the sequences of (abstracted) API calls from the call graph (2). In (3), the sequences of calls are used to build a Markov chain and a feature vector for that app. Finally, classification is performed in (4), classifying the app as benign or malicious.}
 \label{fig:diagram}
\end{figure}

\descr{Building Blocks.} \approach's operation goes through four phases as depicted in
Fig.~\ref{fig:diagram}. First, we extract the call graph from each app by using static analysis (1), then, we obtain the sequences of API calls using all unique nodes after which we abstract each call to class, package, or family (2). Next, we model the behavior of each app by %
constructing Markov chains from the sequences of abstracted API calls for the app (3),
with the transition probabilities used as the feature vector to classify the app as either benign or malware using a machine learning classifier (4).
In the rest of this section, we discuss each of these steps in detail.

\subsection{Call Graph Extraction}\label{sec:extraction}
The first step in \approach is to extract the app's call graph.
We do so by performing static analysis on the app's apk, i.e., the standard Android archive file format  
containing all files, including the Java bytecode, making up the app.
We use a Java optimization and analysis framework, 
Soot~\cite{ValleeRai1999Soot}, to extract call graphs %
and FlowDroid~\cite{Arzt2014flowdroid} to ensure contexts and flows are preserved. %
Specifically, we use FlowDroid, which is based on Soot, to create a dummy main method that serves as the main entry point into the app under analysis (AUA). We do so because Android apps have multiple entry points via which they can be started or invoked. Although apps have an activity launcher, which serves as the main entry point, it is not mandatory that they are implemented (e.g., apps that run as a service), hence, creating a single entry point allows us to reliably traverse the AUA. FlowDroid also lets us model the information flow from sources and sinks using those provided by SuSi~\cite{rasthofer2014machine} as well as callbacks.

To better clarify the different steps involved in our system, we employ throughout this section, a ``running example,'' using a real-world malware sample.
Fig.~\ref{lst:java} lists a class extracted from the decompiled apk of malware disguised as a memory booster app (with package name com.g.o.speed.memboost), which executes commands (rm, chmod, etc.) as root.\footnote{\url{https://www.hackread.com/ghost-push-android-malware/}} 
To ease presentation, we  focus on the portion of the code executed in the try/catch block.
The resulting call graph of the try/catch block is shown in Fig.~\ref{fig:calls}. %
For simplicity, 
we omit calls for object initialization, return types and parameters, as well as
implicit calls in a method. Additional calls that are invoked when getShell(true) is called are not
shown, except for the add() method that is directly called by the program code, as shown in Fig.~\ref{lst:java}.

\begin{figure}[t]
\lstset{ %
breaklines=true,  
showspaces=false,
showstringspaces=false,
columns=flexible,
escapechar=@
}
\begin{lstlisting}[basicstyle=\ttfamily\smallscriptsize,language=Java]
package com.fa.c;

import android.content.Context;
import android.os.Environment;
import android.util.Log;
import com.stericson.RootShell.execution.Command;
import com.stericson.RootShell.execution.Shell;
import com.stericson.RootTools.RootTools;
import java.io.File;

public class RootCommandExecutor {
  public static boolean Execute(Context paramContext) {
    paramContext = new Command(0, new String[] { "cat " + Environment.getExternalStorageDirectory().getAbsolutePath() + File.separator + Utilities.GetWatchDogName(paramContext) + " > /data/" + Utilities.GetWatchDogName(paramContext), "cat " + Environment.getExternalStorageDirectory().getAbsolutePath() + File.separator + Utilities.GetExecName(paramContext) + " > /data/" + Utilities.GetExecName(paramContext), "rm " + Environment.getExternalStorageDirectory().getAbsolutePath() + File.separator + Utilities.GetWatchDogName(paramContext), "rm " + Environment.getExternalStorageDirectory().getAbsolutePath() + File.separator + Utilities.GetExecName(paramContext), "chmod 777 /data/" + Utilities.GetWatchDogName(paramContext), "chmod 777 /data/" + Utilities.GetExecName(paramContext), "/data/" + Utilities.GetWatchDogName(paramContext) + " " + Utilities.GetDeviceInfoCommandLineArgs(paramContext) + " /data/" + Utilities.GetExecName(paramContext) + " " + Environment.getExternalStorageDirectory().getAbsolutePath() + File.separator + Utilities.GetExchangeFileName(paramContext) + " " + Environment.getExternalStorageDirectory().getAbsolutePath() + File.separator + " " + Utilities.GetPhoneNumber(paramContext) });
    try {
      RootTools.getShell(true).add(paramContext);
      return true;
    }
    catch (Exception paramContext) {
     Log.d("CPS", paramContext.getMessage());
    }
    return false;
  }
}
\end{lstlisting}
\vspace{-0.5cm}
\caption{Code from a malicious app (com.g.o.speed.memboost) executing commands as root.}
\label{lst:java}
\end{figure}

\subsection{Sequence Extraction and Abstraction} \label{sec:abstraction}
In its second phase, \approach extracts the sequences of API calls from the call graph and abstracts the calls to one of three modes.

\descr{Sequence Extraction.}  Since \approach uses static analysis, the graph obtained from Soot represents the sequence of functions that are potentially called by the app.
However, each execution of the app could take a specific {\em branch} of the graph and only execute a
subset of the calls.
For instance, when running the code in Fig.~\ref{lst:java} multiple times,
the Execute method could be followed by different calls, e.g., getShell() in the try block only
or getShell() and then getMessage() in the catch block. 

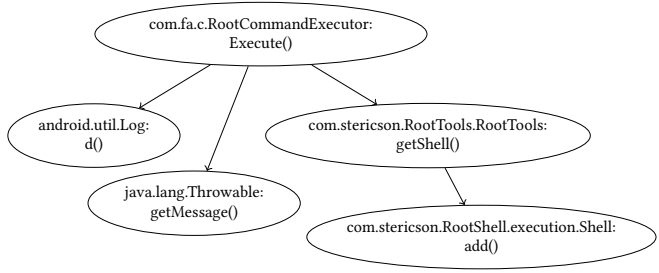
\begin{figure}[t!]
\begin{center}
\resizebox{.625\textwidth}{!}{\scriptsize
\begin{tikzpicture}[every node/.style={draw,ellipse},align=center]
\draw(0,0) node (A0) {com.fa.c.RootCommandExecutor:\\Execute()};
\draw(-2.5,-1.5) node (A1) {android.util.Log:\\ d()};
\draw(2.5,-1.5) node (A2) {\hspace*{-0.15cm}com.stericson.RootTools.RootTools:\hspace*{-0.15cm}\\ getShell()};
\draw(-1,-2.5) node (A3) {java.lang.Throwable:\\ getMessage()};
\draw(3.3,-3) node (A4) {\hspace*{-0.25cm}com.stericson.RootShell.execution.Shell:\hspace*{-0.25cm}\\ add()};
\draw[->] (A0) -- (A1);
\draw[->] (A0) -- (A2);
\draw[->] (A0) -- (A3);
\draw[->] (A2) -- (A4);
\end{tikzpicture}
}
\caption{Call graph of the API calls in the try/catch block of Fig.~\ref{lst:java}. 
(Return types and parameters are omitted to ease presentation).} \label{fig:calls}
\end{center}
\end{figure}

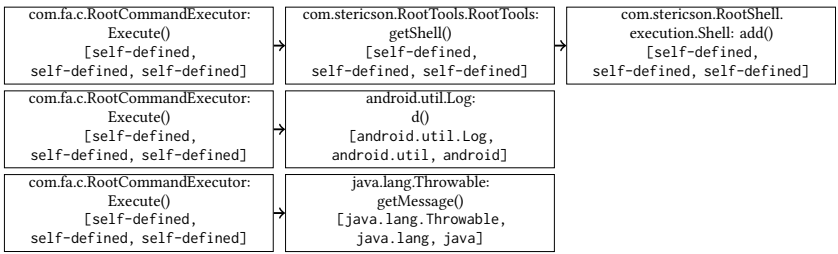
\begin{figure*}[t!]
\begin{center}
\resizebox{0.80\textwidth}{!}{\scriptsize
\begin{tikzpicture}[every node/.style={draw,rectangle},text height=0.1cm,text width=3.90cm,align=center]
\draw(0,0) node (A0) {com.fa.c.RootCommandExecutor: Execute()\\{\tt [self-defined, self-defined, self-defined]}};
\draw(0,-1.25) node (A1) {com.fa.c.RootCommandExecutor:\\Execute()\\{\tt [self-defined, self-defined, self-defined]}};
\draw(0,-2.5) node (A6) {com.fa.c.RootCommandExecutor:\\Execute()\\{\tt [self-defined, self-defined, self-defined]}};
\draw(4.25,0) node (A2) {com.stericson.RootTools.RootTools: getShell()\\{\tt [self-defined, self-defined, self-defined]}};
\draw(8.5,0) node (A3) {com.stericson.RootShell.\\execution.Shell: add()\\{\tt [self-defined, self-defined, self-defined]}};
\draw(4.25,-1.25) node (A4) {android.util.Log:\\ d()\\{\tt [android.util.Log, android.util, android]}};
\draw(4.25,-2.5) node (A5) {java.lang.Throwable:\\getMessage()\\{\tt [java.lang.Throwable, java.lang, java]}};
\draw[->,thick] (A0) -- (A2.west);
\draw[->,thick] (A2) -- (A3.west);
\draw[->,thick] (A1) -- (A4.west);
\draw[->,thick] (A6) -- (A5.west);

\end{tikzpicture}
}
\caption{Sequence of API calls extracted from the call graphs in
Fig.~\ref{fig:calls}, with the corresponding class/package/family abstraction in square brackets.} \label{fig:sequence}
\vspace*{-0.5cm}
\end{center}
\end{figure*}

Thus, in this phase, \approach operates as follows. First, it identifies a set of
entry nodes in the call graph, i.e., nodes with no incoming
edges (for example, the Execute method in the snippet from Fig.~\ref{lst:java} is the entry node if there is no incoming edge from any other call in the app). 
Then, it enumerates the paths reachable from each entry node. The
sets of all paths identified during this phase constitutes the sequences of API calls
which will be used to build a Markov chain behavioral model and
to extract features.
In Fig.~\ref{fig:sequence}, we show the sequence of API calls obtained from the call graph in Fig.~\ref{fig:calls}.
We also report in square brackets, the family, package, and class to
which the call is abstracted.%

\descr{API Call Abstraction.} %
Rather than analyzing raw API calls from the %
sequence of calls, we build \approach to work at a higher level, and operate in one of three modes by abstracting each call to its family, package, or class. 
The intuition is to make \approach resilient to API changes and achieve scalability.
In fact, our experiments presented in Section~\ref{sec:data}, show that, from a
dataset of 44K apps, we extract more than 10 million unique API calls, which, depending on the modeling approach used to model each app, may result in the feature vectors being very sparse.
While package and class are already existing names for these abstraction levels, we use ``family'' to indicate an even higher level of abstraction that does not currently exist. Our use of ``family'' refers to the ``root'' names of the API packages and not to ``malware families,'' since we do not attempt to label each malware sample to its family. When operating in family mode, we abstract an API call to one of the nine Android %
``root'' package names, i.e.,  {\tt android}, {\tt google}, {\tt java}, {\tt javax}, {\tt xml}, {\tt apache}, {\tt junit}, {\tt json}, {\tt dom},  which correspond to the {android.*}, {com.google.*}, {java.*}, {javax.*}, {org.xml.*}, {org.apache.*}, junit.*, org.json, and org.w3c.dom.* packages.
Whereas in package mode, we abstract the call to its package name  %
using the list of Android packages from the documentation\footnote{\url{https://developer.android.com/reference/packages.html}} consisting of 243 packages as of API level 24 (the version as of September 2016), as well as 95 from the Google API.\footnote{\url{https://developers.google.com/android/reference/packages}} In class mode, we abstract each call to its class name using a whitelist of all class names in the Android and Google APIs, which consists respectively, 4,855 and 1116 classes.\footnote{\url{https://developer.android.com/reference/classes.html}}

In all modes, we abstract developer-defined (e.g., com.stericson.roottools) and obfuscated (e.g.
com.fa.a.b.d) API calls respectively, as {\tt self-defined} and {\tt obfuscated}. %
Note that we label an API call as obfuscated if we cannot tell what its class implements, extends, or inherits, due to identifier mangling. %
Overall, there are 11 (9$+$2) families, 340 (243$+$95$+$2) packages, and 5,973 (4,855$+$1,116$+$2) possible classes. %

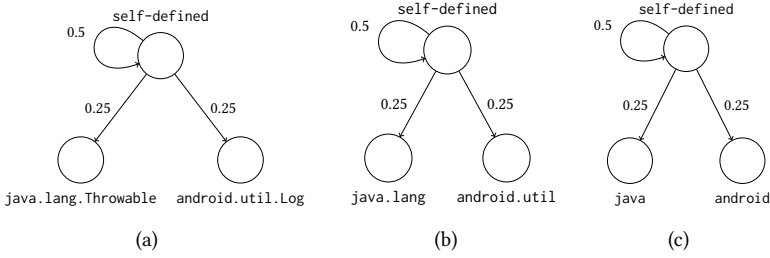
\begin{figure}[t!]
\begin{center}
\subfigure[]{
\resizebox{.305\textwidth}{!}{
\begin{tikzpicture}
\tikzstyle{level 1}=[level distance=1.75cm, sibling distance=3.5cm]
\tikzset{
myroot/.style={label=above:{\tt\strut#1},align=center,anchor=north,draw,circle,minimum size=1cm},
myleaf/.style={label=below:{\tt\strut#1},align=center,anchor=north,draw,circle,minimum size=1cm},
every loop/.style={max distance=200mm,in=-160,out=-220}}
 \node[myroot=self-defined] (q0) {}
    child {node[myleaf=java.lang.Throwable] (q1) {} edge from parent[draw=none]} 
    child {node[myleaf=android.util.Log] (q2) {}  edge from parent[draw=none]}; 
\draw[->] (q0) edge[loop above] () node [midway] {\hspace*{-3.7cm}{0.5}};    
\draw[->] (q0) -- (q1) node [midway=10pt] {\hspace{-1cm}0.25};
\draw[->] (q0) -- (q2) node [midway=10pt] {\hspace{1cm}0.25};
\end{tikzpicture}
}}
\hspace{0.1cm}
\subfigure[]{
\resizebox{.215\textwidth}{!}{
\begin{tikzpicture}
\tikzstyle{level 1}=[level distance=1.75cm, sibling distance=2.5cm]
\tikzset{
myroot/.style={label=above:{\tt\strut#1},align=center,anchor=north,draw,circle,minimum size=1cm},
myleaf/.style={label=below:{\tt\strut#1},align=center,anchor=north,draw,circle,minimum size=1cm},
every loop/.style={max distance=200mm,in=-160,out=-220}}
 \node[myroot=self-defined] (q0) {}
    child {node[myleaf=java.lang] (q1) {} edge from parent[draw=none]} 
    child {node[myleaf=android.util] (q2) {}  edge from parent[draw=none]}; 
\draw[->] (q0) edge[loop above] () node [midway] {\hspace*{-3.7cm}{0.5}};    
\draw[->] (q0) -- (q1) node [midway=10pt] {\hspace{-1cm}0.25};
\draw[->] (q0) -- (q2) node [midway=10pt] {\hspace{1cm}0.25};
\end{tikzpicture}
}}
\hspace{0.1cm}
\subfigure[]{
\resizebox{.178\textwidth}{!}{
\begin{tikzpicture}
\tikzstyle{level 1}=[level distance=1.95cm, sibling distance=2.5cm]
\tikzset{
myroot/.style={label=above:{\tt\strut#1},align=center,anchor=north,draw,circle,minimum size=1cm},
myleaf/.style={label=below:{\tt\strut#1},align=center,anchor=north,draw,circle,minimum size=1cm},
every loop/.style={max distance=200mm,in=-160,out=-220}}
 \node[myroot=self-defined] (q0) {} 
    child {node[myleaf=java] (q1) {}edge from parent[draw=none]} 
    child {node[myleaf=android] (q2) {}edge from parent[draw=none]}; 
\draw[->] (q0) edge[loop above] () node [midway] {\hspace*{-3.7cm}{0.5}};    
\draw[->] (q0) -- (q1) node [midway=10pt] {\hspace{-1cm}0.25};
\draw[->] (q0) -- (q2) node [midway=10pt] {\hspace{1cm}0.25};
\end{tikzpicture}
}}
\end{center}
\vspace{-0.35cm}
\caption{Markov chains originating from the call sequence in Fig.~\ref{fig:sequence} when using classes (a), packages (b) or families (c).}
\label{fig:package}
\vspace*{0.3cm}
\end{figure}

\subsection{Markov-chain Based Modeling} \label{sec:MaCha}
Next, \approach builds feature vectors, used for classification, based on the
Markov chains representing the sequences of abstracted API calls for an app. Before discussing this in detail, %
we first review the basic concepts of Markov chains. 

\descr{Markov Chains.} %
Markov Chains are memoryless models where the probability
of transitioning from a state to another only depends on the current state~\cite{MarCha}. 
They are often represented as a set of nodes, each corresponding to a
different state, and a set of edges connecting one node to another labeled with
the probability of that transition. The sum of all probabilities associated to
all edges from any node (including, if present, an edge going back to the node
itself) is exactly 1. 
The set of possible states of the Markov chain is denoted as $\mathcal{S}$. If $S_{j}$ and $S_{k}$ are two connected states, $P_{jk}$ denotes the probability of transition from $S_{j}$ to $S_{k}$. $P_{jk}$ is given by the number of 
occurrences ($O_{jk}$) of state $S_{k}$ after state $S_{j}$, divided by $O_{ji}$ for all states $i$ in the chain, i.e.,
$P_{jk}=\frac{O_{jk}}{\sum_{i \in \mathcal{S}} O_{ji}}$.

\descr{Building the model.} %
For each app, \approach takes as input the sequence of
abstracted API calls of that app (classes, packages or families, depending on the selected mode of operation), and builds a Markov chain where each class/package/family is a
state and the transitions represent the probability of moving from one state to another. For each Markov chain, state $S_0$ is the entry point from which other calls are made in a sequence.
As an example, Fig.~\ref{fig:package} illustrates the Markov chains built using classes, packages, and families, respectively, from the sequences reported in Fig.~\ref{fig:sequence}.%

We argue that considering single transitions is more robust against attempts to evade detection
by inserting useless API calls in order to deceive signature-based
systems~\cite{kolbitsch2009effective}. 
In fact, \approach considers all possible calls -- i.e., all the branches originating from a node -- in the Markov chain,  so adding calls would not significantly change the probabilities of transitions between nodes (specifically, families, packages, or classes depending on the operational mode) for each app.

\descr{Feature Extraction.}  Next, we use the probabilities of transitioning from one state (abstracted call) to another in the Markov chain as the feature vector of each app. States that are not present in a chain are represented as 0 in the feature vector. The vector derived from the Markov chain depends on the operational mode of \approach. 
With families, there are 11 possible states, thus 121 possible transitions in each chain, while, when abstracting to packages, there are 340 states and 115,600 possible transitions and with classes, there are 5,973 states therefore, 35,676,729 possible transitions.

We also apply Principal Component Analysis (PCA)~\cite{PCA}, which performs feature selection by transforming the feature space into a new space made of components that are linear combinations of the original features. The first component contains as much variance (i.e., amount of information) as possible. The variance %
is given as a percentage of the total amount of information of the original feature space. We apply PCA to the feature set in order to select the principal
components, as PCA  transforms the feature space into a smaller one where the variance is represented with as few components as possible, thus considerably
reducing computation/memory complexity. %
Also, PCA could reduce overfitting by only building the model from the principal components of the features in our dataset which may in turn, improve the accuracy of the classification. %

\subsection{Classification}\label{sec:classification}
The last step is to perform classification, i.e., labeling apps as either benign or malware. To this end, we test \approach using different classification algorithms: Random Forests, %
1-Nearest Neighbor (1-NN), %
3-Nearest Neighbor (3-NN), %
and Support Vector Machines (SVM). %
Note that since both accuracy and speed are worse with SVM, 
we omit results obtained with it. On average, the F-Measure with SVM using Radial Basis Functions (RBF) is 0.09 lower than with Random Forests, and %
it is 5 times slower to train in family mode (which has a much smaller feature space) than 3-Nearest Neighbors (the slowest among the other classification methods).

Each model is trained using the feature vector obtained from the apps in a training sample. Results are presented and discussed in Section \ref{sec:markoveval}, and have been validated by using 10-fold cross validation.
Note that due to the different number of features used in different modes, we use two %
distinct configurations for the Random Forests algorithm. Specifically, when abstracting to families, we use 51 trees with maximum depth 8, while, with classes and packages, we use 101 trees of maximum depth 64. To tune Random Forests we follow the methodology applied in~\cite{RandTune}.%

\begin{table*}[t]
\centering
\footnotesize
\begin{tabular}{|l|l|ll|rrr|}
\multicolumn{7}{c}{}\\
\hline
{\bf Category} & {\bf Name} & \multicolumn{2}{l|}{\bf Date Range}  & {\bf \#Samples} & {\bf \#Samples}  & {\bf \#Samples}\\
& & & & & {\bf (API Calls)} & {\bf (Call Graph)}\\
\hline
\multirow{2}{*}{\em Benign} & {\tt oldbenign} & Apr 2013\hspace*{-0.15cm}&-- Nov 2013  & 5,879 & 5,837 & 5,572 \\
 & {\tt newbenign} & Mar 2016\hspace*{-0.15cm} &-- Mar 2016 & 2,568 & 2,565 & 2,465\\
 \hline
 \multicolumn{4}{|r}{\em Total Benign:} & \multicolumn{1}{r}{\em 8,447} & \multicolumn{1}{r}{\em 8,402} & \multicolumn{1}{r|}{\em 8,037}\\[1ex]
\hline
\multirow{5}{*}{\em Malware} & {\tt drebin} & Oct 2010\hspace*{-0.15cm} &-- Aug 2012 & 5,560 & 5,546 & 5,512\\
 & {\tt 2013} &  Jan 2013\hspace*{-0.15cm} &-- Jun 2013 & 6,228 & 6,146 & 6,091\\
& {\tt 2014} & Jun 2013\hspace*{-0.15cm} &-- Mar 2014 &  15,417 & 14,866 & 13,804\\
 & {\tt 2015} & Jan 2015\hspace*{-0.15cm} &-- Jun 2015  & 5,314 & 5,161 & 4,451\\
 & {\tt 2016} & Jan 2016\hspace*{-0.15cm} &-- May 2016 & 2,974 & 2,802 & 2,555 \\
 \hline
  \multicolumn{4}{|r}{\em Total Malware:} & \multicolumn{1}{r}{\em 35,493} & \multicolumn{1}{r}{\em 34,521} & \multicolumn{1}{r|}{\em 32,413}\\[0.5ex]
  \hline
\end{tabular}
\caption{Overview of the datasets used in our experiments.} 
\label{table:dataset}
\vspace{-0.15cm}
\end{table*}

\section{Dataset}\label{sec:data}

In this section, we introduce the datasets used in the evaluation of \approach (presented later in Section~\ref{sec:markoveval}),
which include 43,940 apk files, specifically, 8,447 benign and 35,493 malware samples.
We include a mix of older and newer apps, ranging from October 2010 to May 2016, 
as we aim to verify that \approach is robust to changes in Android malware samples as well as APIs.
Also, to the best of our knowledge,
our evaluation is done on one of the largest Android malware datasets used in a research paper. When evaluating \approach, we use one set of malicious samples (e.g., {\tt drebin}) and a benign set (e.g., {\tt oldbenign}), thus experimenting %
with a balanced dataset among the two classes, as shown in Table \ref{table:dataset}.

\descr{Benign Samples.} Our benign datasets consist of two sets of samples: (1) one, which we denote as {\tt oldbenign}, includes 5,879 apps collected by PlayDrone~\cite{viennot2014measurement} between April and November 2013, and published on the Internet Archive\footnote{\url{https://archive.org/details/playdrone-apk-e8}} on August 7, 2014; and (2) another, {\tt newbenign}, 
obtained by downloading the top 100 apps in each of the 29 categories on the Google Play store as of March 7, 2016, using the googleplay-api tool.\footnote{\url{https://github.com/egirault/googleplay-api}} Due to errors encountered while downloading some apps, we have actually obtained 2,843 out of 2,900 apps. Note that 275 of these belong to more than one category, therefore, the \texttt{newbenign} dataset ultimately includes 2,568 unique apps.

\descr{Android Malware Samples.} The set of malware samples includes apps that were used to test
{\sc Drebin}~\cite{arp2014drebin}, dating back to October 2010 -- August 2012
(5,560), which we denote as {\tt drebin}, as well as more recent ones that have
been uploaded on the VirusShare\footnote{\url{https://virusshare.com/}} site over the years. Specifically, we gather from VirusShare, respectively, 6,228, 15,417, 5,314, and 2,974 samples from {\tt 2013}, {\tt 2014}, {\tt 2015}, and {\tt 2016}. We consider each of these datasets separately for our analysis.

\descr{API Calls.} For each app, we extract all API calls, using Androguard\footnote{\url{https://github.com/androguard/androguard}}, since, as explained in Section~\ref{sec:compare}, these constitute the features used by \droid~\cite{Aafer2013DroidAPIMiner} (against which we compare our system) %
as well as a variant of \approach that is based on frequency analysis (see Section~\ref{sec:evaluation}). 
 Due to Androguard failing to decompress some of the apks, bad CRC-32 redundancy checks, and errors during unpacking, we are not able to extract the API calls for all the samples, but
only for 40,923 (8,402 benign, 34,521 malware) out of the 43,940 apps in our datasets.

\descr{Call Graphs.} To extract the call graph of each apk, we use Soot and FlowDroid. 
Note that for some of the larger apks, Soot requires a non-negligible amount of memory to extract the call graph, so 
we allocate 16GB of RAM to the Java VM heap space. %
We find that for 
2,472 (364 benign + 2,108 malware) samples, Soot is not able to complete the
extraction due to it failing to apply the {\tt jb} phase as well as reporting an
error in opening some zip files (i.e., the apk).  The {\tt jb} phase is used by Soot to transform Java bytecode into jimple intermediate representation (the primary IR of Soot) for optimization purposes.
Therefore, we exclude these apps in our evaluation and discuss this limitation further in Section~\ref{sec:limits}. %
In Table~\ref{table:dataset}, we provide a summary of our seven datasets, reporting the total number of samples per dataset,
as well as those for which we are able to extract the API calls (second-to-last column) and the call graphs (last column). 

\begin{figure}[t!]
\centering
\hspace*{-0.3cm}
\subfigure[\label{fig:NumCalls}API calls]
{\includegraphics[width=0.34\textwidth]{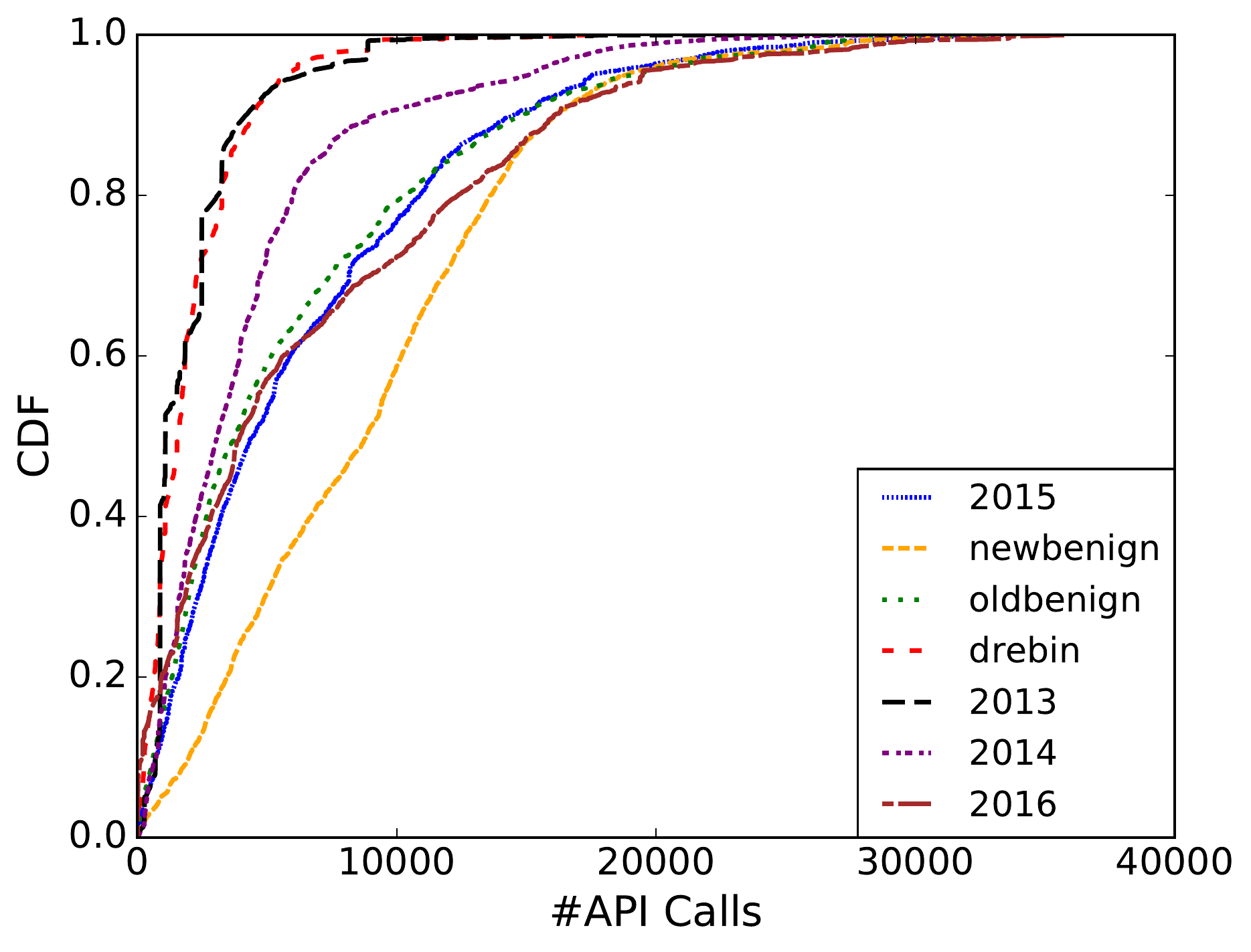}}
\hspace*{-0.2cm}
\subfigure[\label{fig:FamAndro}{\tt android}]
{\includegraphics[width=0.34\textwidth]{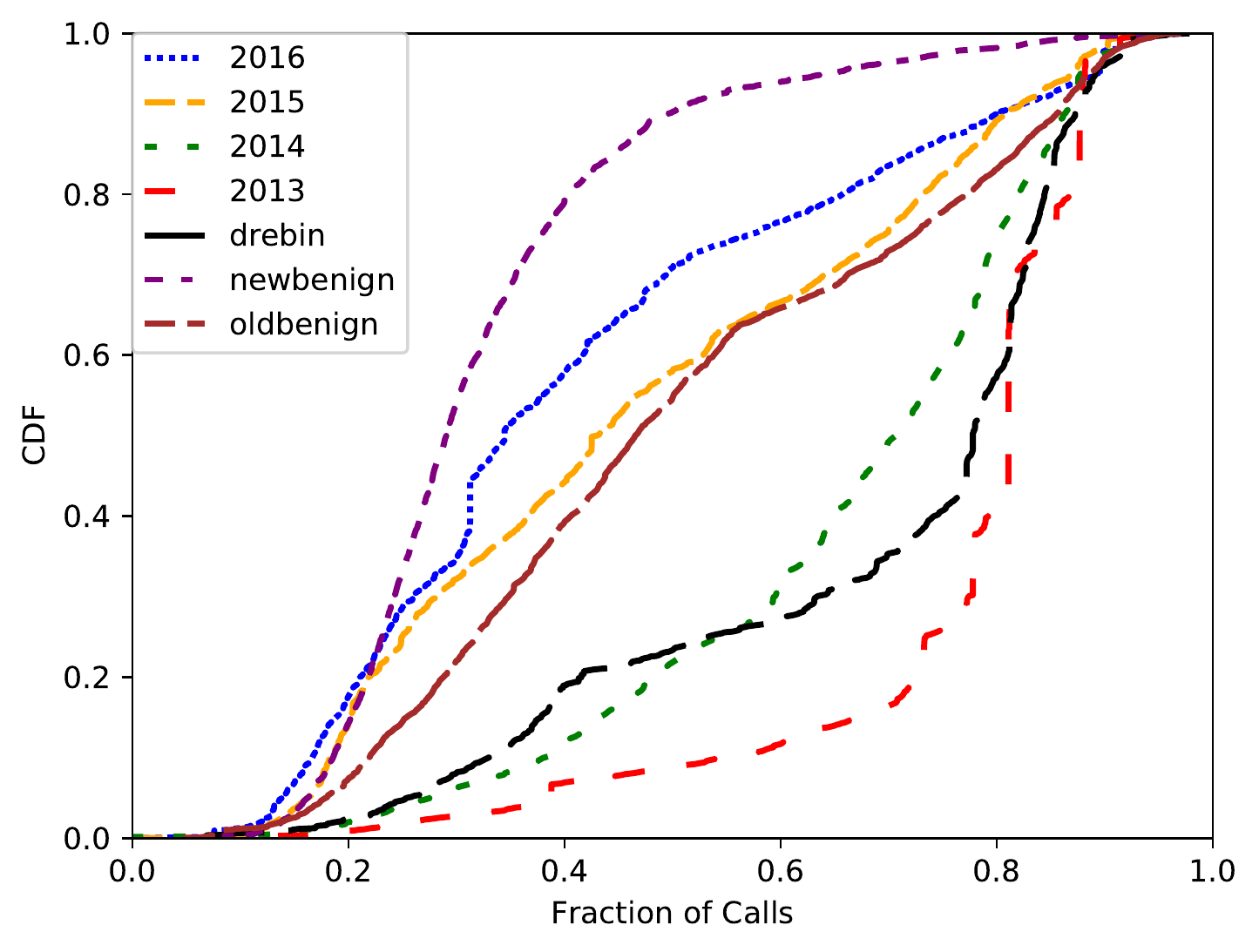}}
\hspace*{-0.2cm}
\subfigure[\label{fig:FamGoogle}{\tt google}]
{\includegraphics[width=0.34\textwidth]{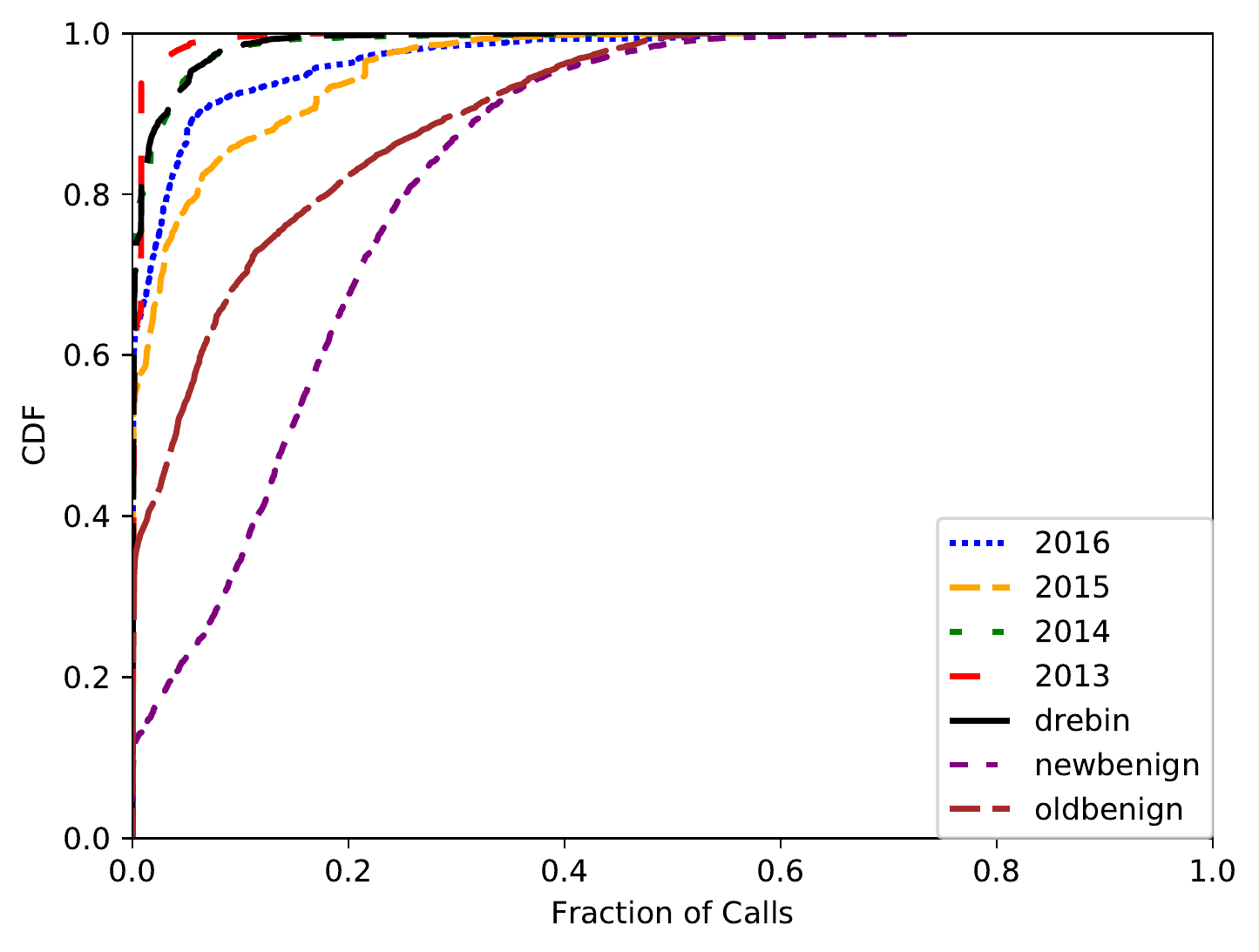}}
\vspace{-0.4cm}
\caption{CDFs of the number of API calls in different apps in each dataset (a), and of the percentage of {\tt android} (b) and {\tt google} (c) family calls.}
\label{fig:androidapi}
\vspace{0.2cm}
\end{figure}

\descr{Dataset Characterization}. 
Aiming to shed light on the evolution of API calls in Android apps,
we also performed some measurements over our datasets. %
In Fig.~\ref{fig:NumCalls}, we plot the Cumulative Distribution Function
(CDF) of the number of unique API calls in the apps in different datasets, highlighting that newer apps, both
benign and malicious, %
use more API calls overall than older apps. This
indicates that as time goes by, Android apps become more complex.
When looking at the fraction of API calls belonging to specific families, we
discover some interesting aspects of Android apps developed in different years.
In particular, we notice that API calls belonging to the {\tt android} family become less
prominent as time passes (Fig.~\ref{fig:FamAndro}), both in benign and
malicious datasets, while {\tt google}
calls become more common in newer apps (Fig.~\ref{fig:FamGoogle}).
In general, we conclude that benign and malicious apps show the same
evolutionary trends over the years. Malware, however, appears to reach the same
characteristics (in terms of level of complexity and fraction of API calls from
certain families) as legitimate apps with a few years of delay.

\section{\approach Evaluation}
\label{sec:markoveval}
We now present an experimental evaluation of \approach when it operates in family or package mode. 
Later in Section~\ref{sec:classeval}, we evaluate it in class mode. We use the datasets summarized in Table~\ref{table:dataset}, and evaluate \approach, as per 
(1) its accuracy on benign and malicious samples developed around the same time; 
and (2) its robustness to the evolution of malware as well as of the Android framework by using older datasets for training and newer ones for testing and vice-versa.

\subsection{Experimental Settings}
\begin{figure}[t]
\centering
\subfigure[\label{fig:FM10fam}{family mode}]
{\includegraphics[width=0.45\textwidth]{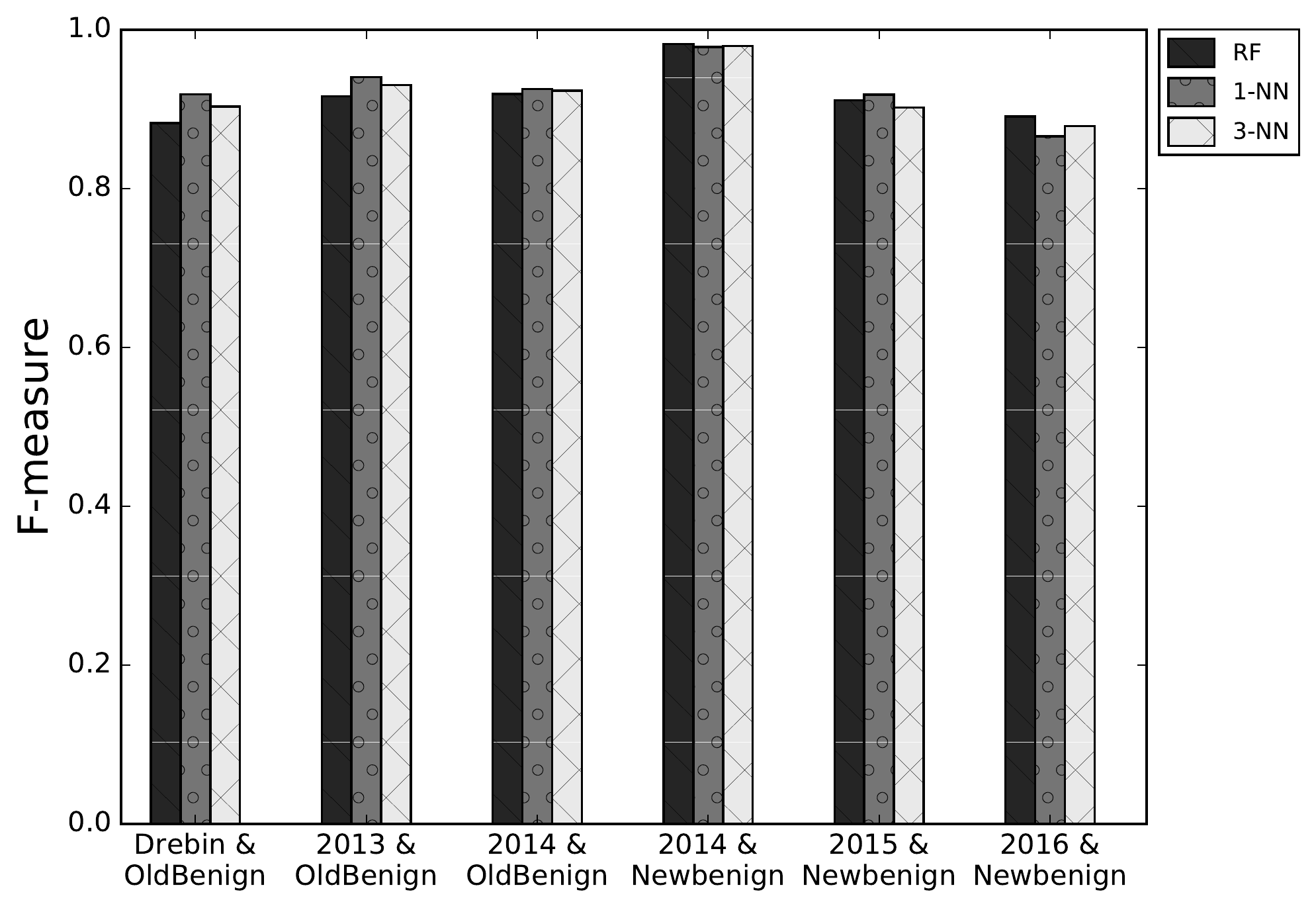}}
\subfigure[\label{fig:FM10pac}{package mode}]
{\includegraphics[width=0.45\textwidth]{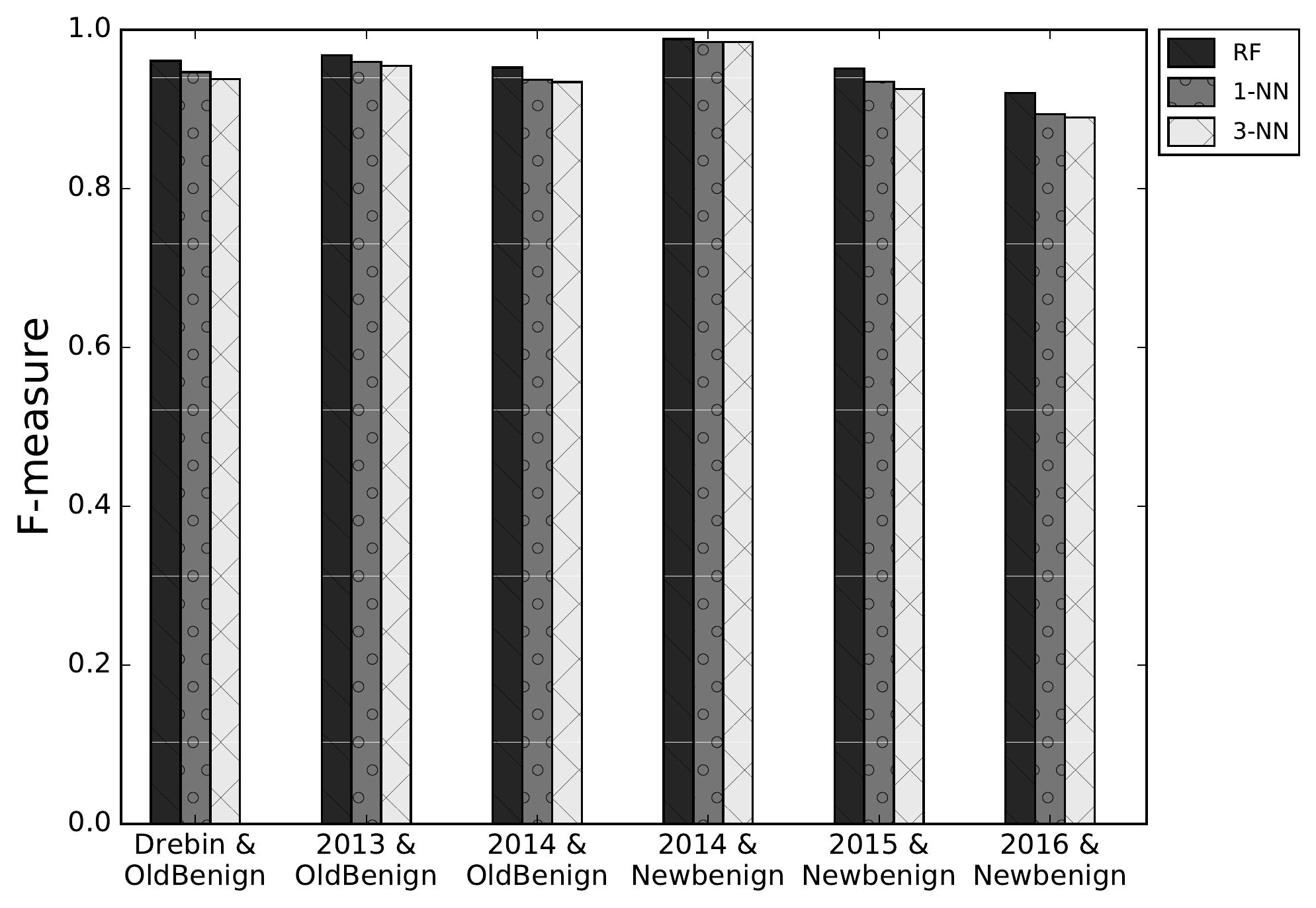}}
\vspace{-0.4cm}
\caption{F-measure of \approach classification with datasets from the same year using three different classifiers.} 
\label{fig:mamabarsame}
\end{figure}

To assess the accuracy of the classification, we use the standard F-measure metric, 
calculated as 
$F=2\cdot(\mbox{Precision}\cdot\mbox{Recall})/(\mbox{Precision}+\mbox{Recall})$,
where Precision $=$ TP$/($TP$+$FP$)$ and 
Recall $=$ TP$/($TP$+$FN$)$.
TP denotes the number of samples correctly classified as malicious, while
FP an FN indicate, respectively, the number of samples mistakenly identified as malicious and benign. 

Note that all our experiments perform 10-fold cross validation using at
least one malicious and one benign dataset from Table~\ref{table:dataset}. In
other words, after merging the datasets, the resulting set is shuffled and
divided into ten equal-size random subsets. Classification is then performed ten
times using nine subsets for training and one 
for testing, and results are
averaged out over the ten experiments.

When implementing \approach in family mode, we exclude %
{\tt json} and {\tt dom}  families because they are almost never used across all our datasets, and {\tt junit}, which is primarily used for testing.
In package mode, in order to avoid incorrect abstraction when {\tt self-defined} APIs have ``android'' in the name, we split the {\tt android} package into its  two classes, i.e., {\tt android.R} and {\tt android.Manifest}.
Therefore, in family mode, there are 8 possible states, thus 64 features, whereas in package mode, we have 341 states and 116,281 features (cf.~Section~\ref{sec:MaCha}). %

\begin{table*}[t]
\centering
\scriptsize
\setlength{\tabcolsep}{3pt}
\resizebox{0.99\linewidth}{!}{
\begin{tabular}{|l|p{0.75cm}p{0.75cm}r|rrr|p{0.75cm}p{0.75cm}r|rrr|}
\hline
\multirow{ 2}{*}{\backslashbox{\bf Dataset}{\bf Mode}}  & \multicolumn{12}{c|}{[Precision, Recall, $\textbf{F}$-measure]}   \\
\cline{2-13}
 & \multicolumn{3}{c|}{Family} & \multicolumn{3}{c|}{Family (PCA)} & \multicolumn{3}{c|}{Package} & \multicolumn{3}{c|}{Package (PCA)} \\
\hline
{\texttt{drebin}, {\tt oldbenign}} &  0.82 & 0.95 & 0.88 & 0.84 & 0.92 & 0.88 & 0.95 & 0.97 & 0.96 & 0.94 & 0.95 & 0.94 \\
\hline
{\texttt{2013}, {\tt oldbenign}} & 0.91 & 0.93 & 0.92 & 0.93 & 0.90 & 0.92 & 0.98 &  0.95  & 0.97 & 0.97 & 0.95 &  0.96 \\
\hline
{\texttt{2014}, {\tt oldbenign}} & 0.88 & 0.96 & 0.92 & 0.87 & 0.94 & 0.90 & 0.93 &  0.97 & 0.95 & 0.92 &  0.96 &  0.94 \\
\hline
{\texttt{2014}, {\tt newbenign}} & 0.97 &  0.99 &  0.98 & 0.96 & 0.99 & 0.97 & 0.98 &  1.00 & 0.99 & 0.97 & 1.00 & 0.99 \\
\hline
{\texttt{2015}, {\tt newbenign}} & 0.89 & 0.93 & 0.91 & 0.87 & 0.93 & 0.90 & 0.93 & 0.98 & 0.95 & 0.91 & 0.97 & 0.94 \\
\hline
{\texttt{2016}, {\tt newbenign}} & 0.87 & 0.91 & 0.89 & 0.86 & 0.88 & 0.87 & 0.92 & 0.92 & 0.92 & 0.88 & 0.89 & 0.89 \\
\hline
\end{tabular}
}
\caption{Precision, Recall, and F-measure obtained by \approach when trained and tested with dataset from the same year, %
with and without PCA.}
\label{table:overallresults}
\vspace{-0.25cm}
\vspace{0.25cm}
\end{table*}

\subsection{\approach's Performance (Family and Package Mode)}\label{sec:mama}
We start by evaluating the performance of \approach when it is trained and tested on dataset from the same year. %

In Fig.~\ref{fig:mamabarsame}, we plot the F-measure achieved by \approach in family and package modes using datasets from the same year for training and testing and the three different classifiers. %
As already discussed in Section~\ref{sec:MaCha}, we apply PCA %
as it allows us transform a large feature space into a smaller one. 
When operating in package mode, PCA could be particularly beneficial to reduce computation and memory complexity, since \approach originally has to operate over 116,281 features. Hence, in Table~\ref{table:overallresults} we report the Precision, Recall, and F-measure achieved by \approach in both modes with and without the application of PCA using Random Forest classifier. We report the results for Random Forest only because it outperforms both 1-NN and 3-NN (Fig.~\ref{fig:mamabarsame}) while also being very fast.
In package mode, we find that only 67\% of the variance is taken into account by the 10 most important PCA components, and in family mode, at least 91\% of the variance is included by the 10 PCA Components. %
As shown in Table~\ref{table:overallresults}, the F-measure using PCA is only slightly lower (up to 3\%) than using the full feature set. In general, \approach performs better in package mode in all datasets with F-measure ranging from 0.92 -- 0.99 compared to 0.88 -- 0.98 in family mode. This is as a result of the increased granularity which enables \approach identify more differences between benign and malicious apps. On the other hand, however, this likely reduces the efficiency of the system, as many of the states derived from the abstraction are used only a few times. The differences in time performance between the two modes are analyzed in details in Section~\ref{sec:runtime}.

\begin{figure}[t]
\centering
\subfigure[\label{fig:FMFutfam}{family mode}]
{\includegraphics[width=0.38\textwidth]{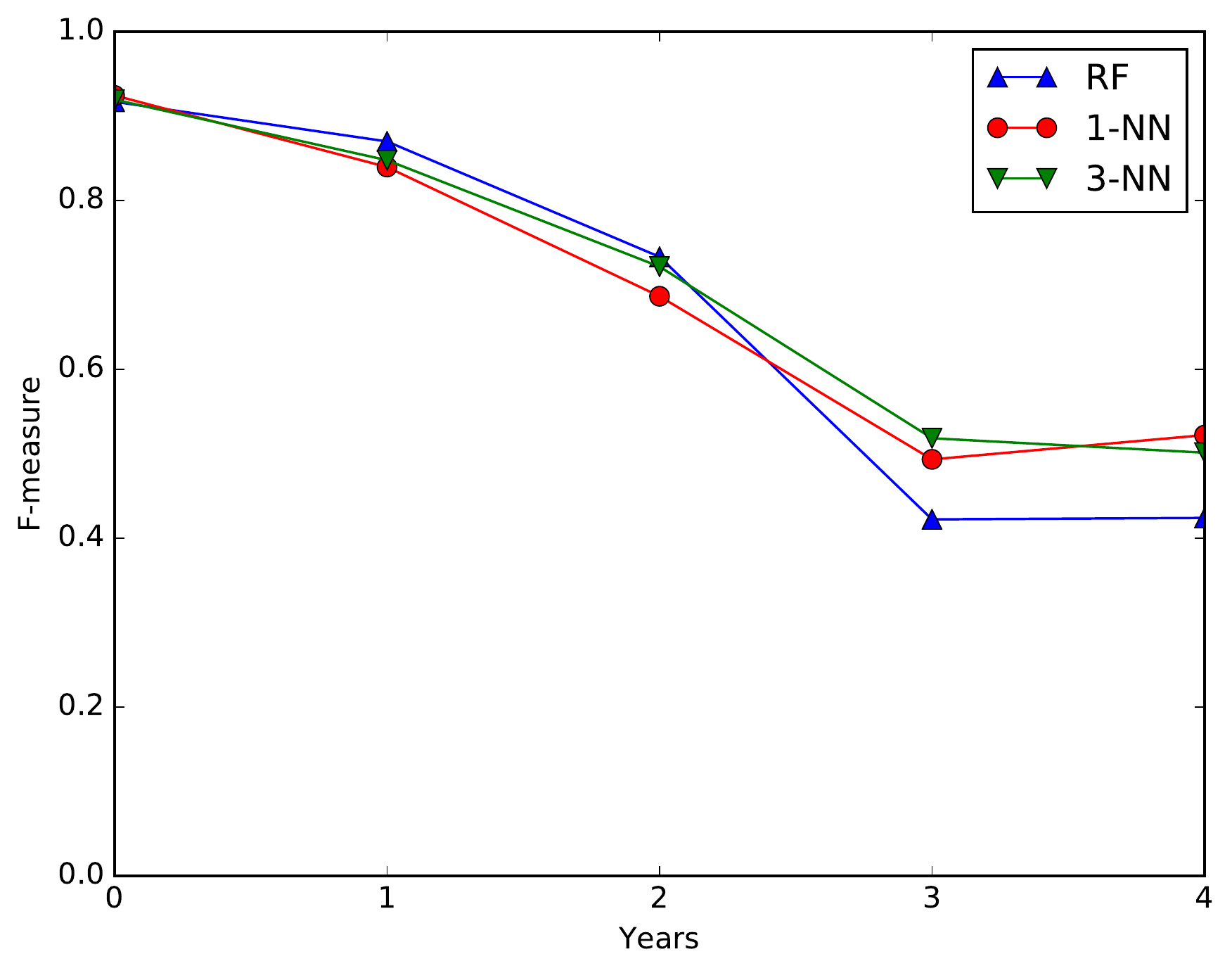}}
\hspace{0.5cm}
\subfigure[\label{fig:FMFutpac}{package mode}]
{\includegraphics[width=0.38\textwidth]{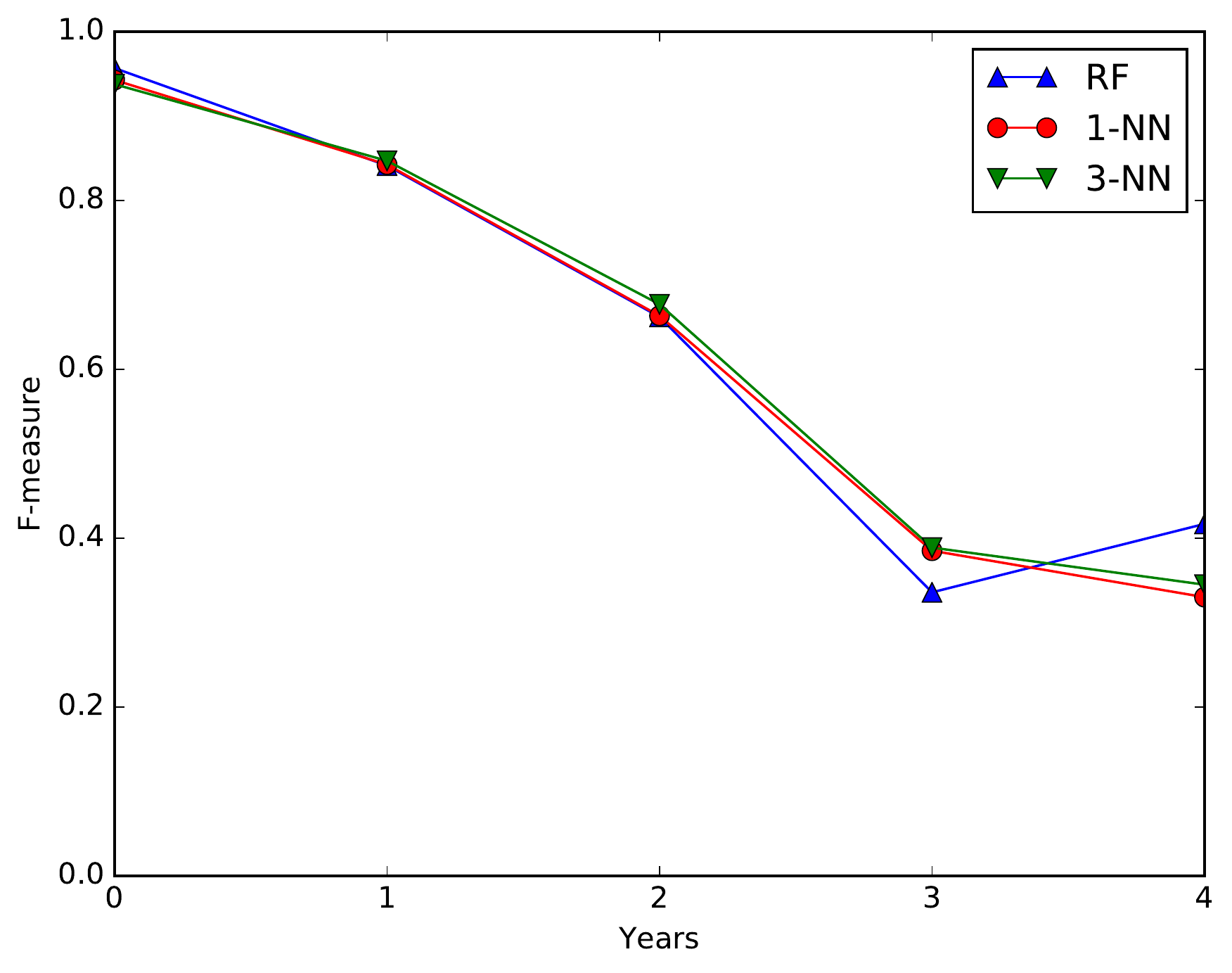}}
\vspace{-0.5cm}
\caption{F-measure achieved by \approach using {\em older} samples for training and {\em newer} samples for testing.} %
\label{fig:PmarkovF} 
\end{figure}

\subsection{Detection Over Time}\label{sec:detectiontime}
As Android evolves over the years, so do the characteristics 
of both benign and malicious apps.  Such evolution must be taken into account when evaluating Android malware detection systems, since their accuracy might significantly be affected as newer APIs are released and/or as malicious developers modify their strategies in order to avoid detection. Evaluating this aspect constitutes one of our research questions, and one of the reasons why our datasets span across multiple years (2010--2016).

Recall that \approach relies on the sequence of API calls extracted from the
call graphs and abstracted to either the package or the family level. Therefore, it is less susceptible to changes in the Android API
than other classification systems such as \droid~\cite{Aafer2013DroidAPIMiner} 
and {\sc Drebin}~\cite{arp2014drebin}. Since these rely on the
use or the frequency, of certain API calls to classify malware vs benign samples, they need to be
retrained following new API releases. On the contrary, retraining is not needed as often with \approach, since families and packages represent
more abstract functionalities that change less over time.
Consider, for instance, the {\tt android.os.health} package released in API level 24; it contains a set of
classes that helps developers track and monitor system
resources.\footnote{\url{https://developer.android.com/reference/android/os/health/package-summary.html}} Classification systems built before this release -- as in the case of \droid~\cite{Aafer2013DroidAPIMiner} (released in 2013, when Android API was up to level 20) --
need to be retrained if this package is more frequently used by malicious apps than benign apps, while \approach only needs to add a new state to its Markov chain when operating in package mode, while no additional state is required when operating in family mode.  

\descr{Older training, newer testing.} To verify this hypothesis, we test \approach using older samples as training sets and newer ones as test sets.
Fig.~\ref{fig:FMFutfam} reports the average F-measure of
the classification in this setting, with \approach operating in
family mode. The x-axis reports the difference in years between the training and testing malware
dataset. We obtain 0.86 F-measure when we classify
apps one year older than the samples on which we train. Classification is still relatively accurate, at 0.75, even after two years.
Then, from
Fig.~\ref{fig:FMFutpac}, we observe that the average F-measure does not significantly change when operating in package
mode. Both modes of operations are affected by one particular condition, already discussed in Section
\ref{sec:data}: in our models, benign datasets seem to ``precede'' malicious ones by 1--2 years in the way they use certain API calls. 
As a result, we notice a drop in accuracy when classifying future samples and using {\tt drebin} (with samples from 2010 to 2012) or {\tt 2013} as the malicious training set and {\tt oldbenign} (late 2013/early 2014) as the benign training set.
More specifically, we observe that \approach correctly detects benign apps,
while it starts missing true positives and increasing false negatives --- i.e., achieving lower Recall.

\begin{figure}[t]
\centering
\subfigure[\label{fig:FMPastfam}{family mode}]
{\includegraphics[width=0.38\textwidth]{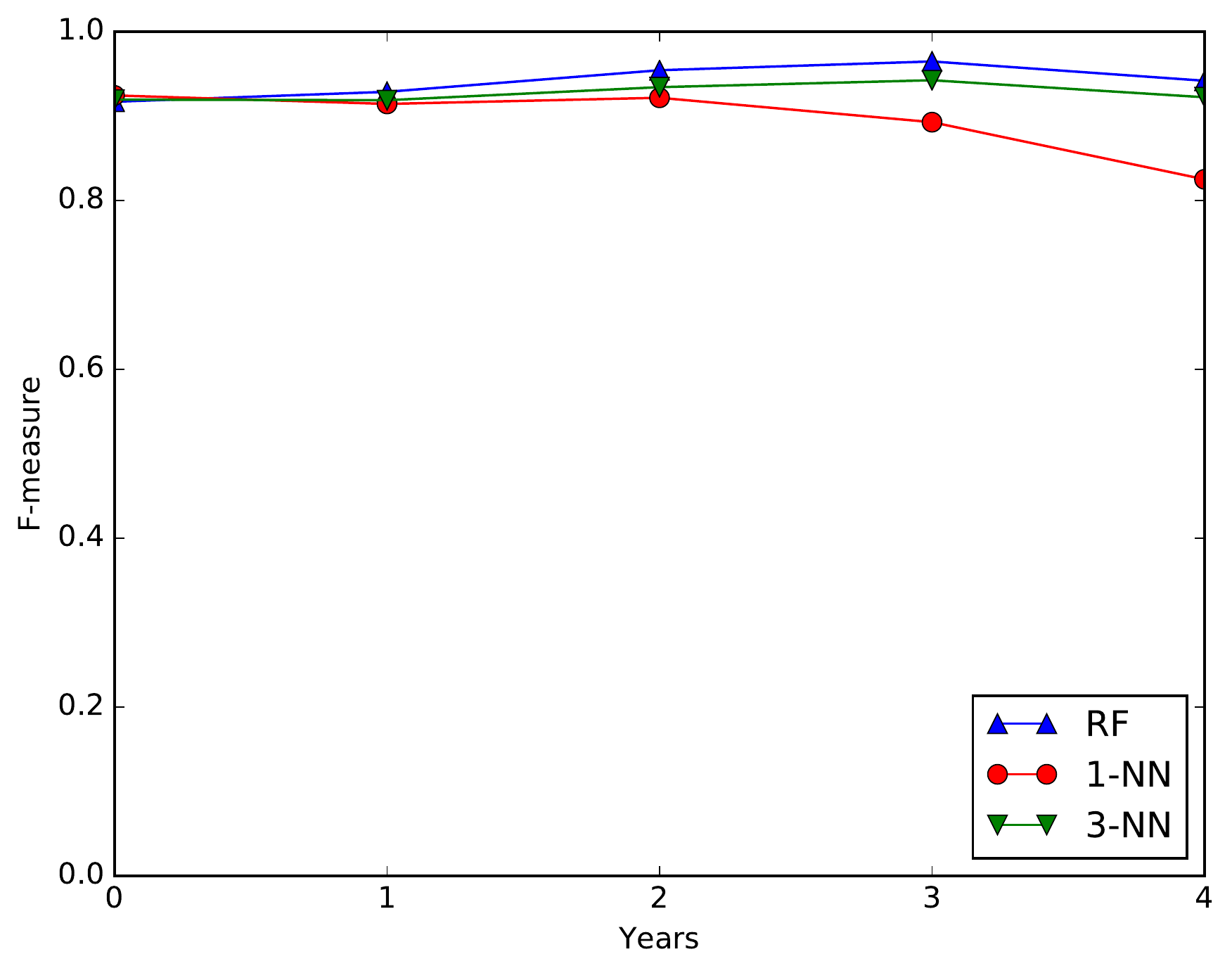}}
\hspace{0.5cm}
\subfigure[\label{fig:FMPastpac}{package mode}]
{\includegraphics[width=0.38\textwidth]{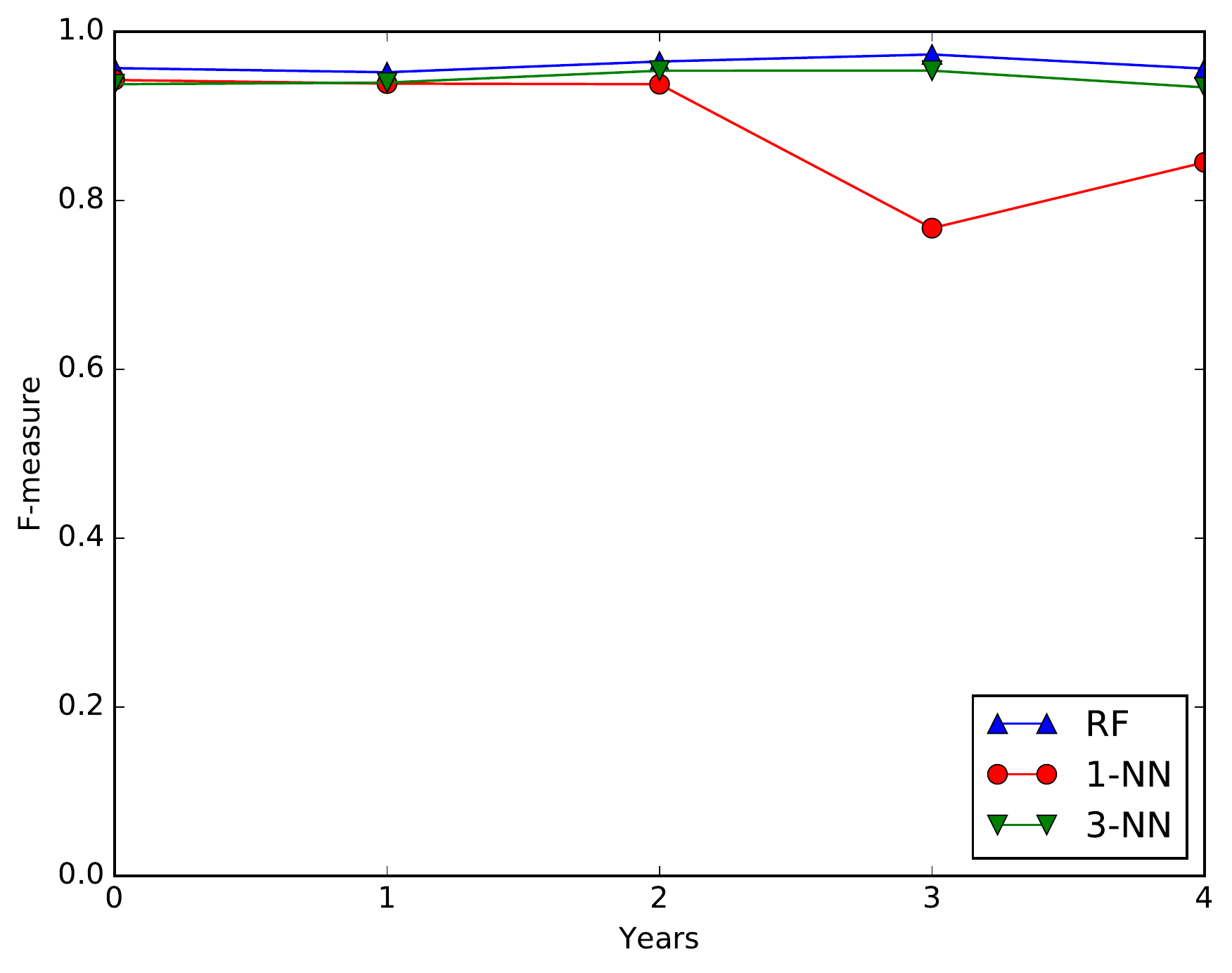}}
\vspace{-0.4cm}
\caption{F-measure achieved by \approach using {\em newer} samples for training and {\em older} samples for testing.} %
\label{fig:FmarkovP}
\end{figure}

\descr{Newer training, older testing.}  We also set to verify whether older malware samples can still be detected
by the system---if not, this would obviously become vulnerable to older (and possibly popular) attacks. 
Therefore, we also perform the ``opposite'' experiment, i.e., training \approach
with newer benign (March 2016) and malware (early 2014 to mid 2016) datasets, and checking whether it is able to detect malware developed years
before. Specifically, Fig.~\ref{fig:FMPastfam} and~\ref{fig:FMPastpac} report results when training
\approach with samples from a given year, and testing it with others that are up to 4 years older:  
\approach retains similar F-measure scores over the years. Specifically, in family mode, 
it varies from 0.93 to 0.96, whereas in package mode, from 0.95 to 0.97 with the oldest samples. 

Note that as \approach does general malware classification and not targeted malware family classification, we believe that its detection performance over time is not affected by the family of malware in our datasets. For example, the {\tt drebin} dataset comprises about 1,048 malware families~\cite{arp2014drebin} which may not all be in the, e.g., {\tt 2016} dataset. As a result, \approach has no prior knowledge of all the malware samples in the testing samples when it does classification using older samples for training.

\subsection{Case Studies of False Positives and Negatives}
The experiment analysis presented above show that \approach detects Android
malware with high accuracy. As in any detection system, however, the system makes a small number of incorrect classifications,
incurring some false positives and false negatives.
Next, we discuss a few case studies aiming to better understand these misclassifications.
We focus on the experiments with newer datasets, i.e., \texttt{2016} and \texttt{newbenign}.

\begin{figure}
\centering
\includegraphics[scale=0.43]{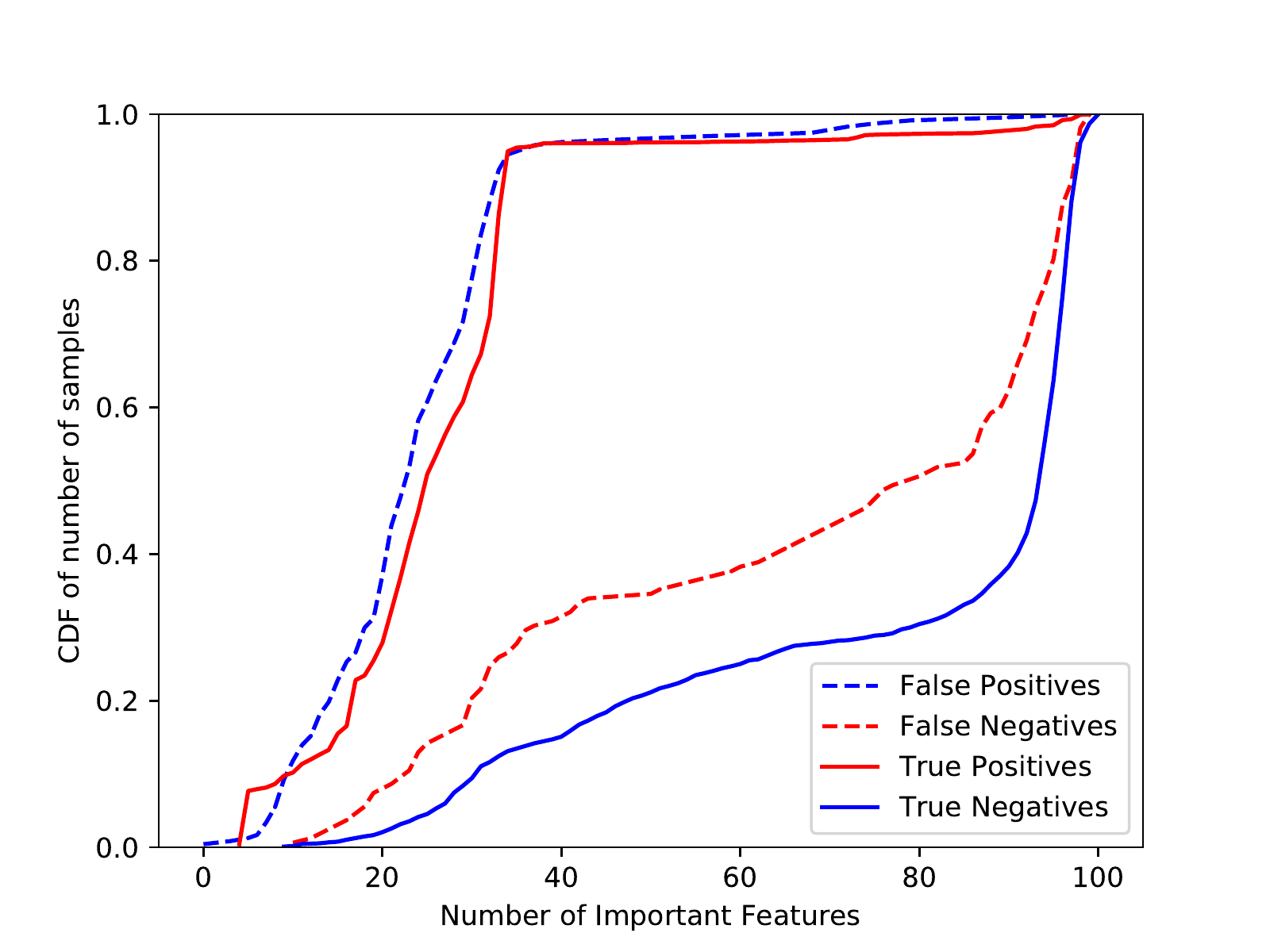}
\vspace{-0.4cm}
\caption{CDF of the number of important features in each classification type.}
\label{fig:fpfn}
\end{figure}

\descr{False Positives.} First, we analyze the manifest of 164 apps
mistakenly detected as malware by \approach, finding that most of them use ``dangerous'' 
permissions~\cite{andriotis2016permissions}. In particular, 67\% write to external storage, 32\% read the phone state, and 21\% access the device's fine location. We further analyzed %
apps (5\%) that use the READ\_SMS and SEND\_SMS permissions, i.e., even though they
are not SMS-related apps, they can read and send SMSs
as part of the services they provide to users. In particular, a {\em ``in case of emergency''} app is
able to send messages to several contacts from its database (possibly added by the user), which
is a typical behavior of Android malware in our dataset, ultimately leading \approach
to flag it as malicious.
As there are sometimes legitimate reasons for benign apps to use permissions considered to be dangerous, we also analyze the false positives using a second approach. Specifically, we examine the average number of the 100 most important features used by \approach to distinguish malware from benign apps that are present in the false positive samples when operating in package mode. We select the top 100 features as it represents no more than about 4.5\% of the features (there are 2202 features in the \texttt{2016} and \texttt{newbenign} datasets). As shown in Fig.~\ref{fig:fpfn}, we find that for 90\% of the false positives, there are only, on average, 33 of the most important features present in their feature vectors which are similar to the behavior observed in the true positives (34/100). \approach misclassifies these samples because they exhibit similar behavior to the true positives and these samples could be further manually analyzed to ascertain maliciousness.

\descr{False Negatives.} We also check 114 malware samples missed
  by \approach when operating in family mode, using VirusTotal.\footnote{\url{https://www.virustotal.com}} We find that 18\%
  of the false negatives are actually not classified as malware by any of the antivirus engines used
  by VirusTotal, suggesting that these are actually legitimate apps  mistakenly included in
  the VirusShare dataset. 45\% of \approach's false negatives are {\em adware}, typically,
  repackaged apps in which the advertisement library has been substituted with a third-party one,
  which creates a monetary profit for the developers. Since they are not performing any clear malicious activity,
  \approach is unable to identify them as malware. 
  Finally, we find that 16\% of the false negatives reported by \approach are samples sending text messages or
starting calls to premium services. We also do a similar analysis of false negatives
when abstracting to packages (74 samples), with similar results: there a few more adware samples (53\%), but similar
percentages for potentially benign apps (15\%) and samples sending SMSs or placing calls (11\%). Similarly, we also investigate the false negatives further, using the 100 most important features of \approach when operating in package mode. As shown in Fig.~\ref{fig:fpfn}, we find that for 90\% of the false negatives, they have on average, 97 of the 100 features similar to the true negatives (98/100).

\subsection{\approach vs \droid}\label{sec:compare}
We also compare the performance of \approach to previous work using API features for Android malware classification. Specifically, to \droid~\cite{Aafer2013DroidAPIMiner} because: (i) it uses API calls and its parameters to perform classification; %
(ii) it reports high true positive rate (up to 97.8\%) on almost 4K malware samples obtained from McAfee and {\sc Genome}~\cite{Zhou2012dissecting}, and 16K benign samples; and (iii) its source code has been made available to us by the authors. 

In \droid, permissions that are requested more frequently
by malware samples than by benign apps are used to perform a baseline classification. 
Then, the system also applies frequency analysis on the list of API calls after removing API calls from ad libraries, using the 169 most frequent 
API calls in the malware samples (occurring at least 6\% more in malware than benign samples). %
Finally, data flow analysis is applied on the API calls that are frequent in both
benign and malicious samples, but do not occur by at least 6\% more in the malware set. Using the top 60 parameters, the 169 most frequent calls change, and the authors report a Precision of 97.8\%. 

After obtaining \droid's source code from the authors, as well as a list of packages (i.e., ad libraries) used for feature refinement,
we re-implement the system by modifying the code in order to reflect recent changes in Androguard (used by \droid for API call extraction),
extract the API calls for all apps in the datasets listed in Table~\ref{table:dataset}, and perform a frequency analysis on the calls.  
Recall that Androguard fails to extract calls for about 2\% (1,017)
of apps, thus \droid is evaluated over the samples in the second-to-last column of Table~\ref{table:dataset}. 
We also implement classification, which is missing from the code
provided by the authors, using k-NN (with k$=$3) since it achieves the best results according to the paper. We use 2/3 of the dataset for training and 1/3 for testing as implemented by the authors.%

In Table~\ref{table:droidapiresults}, we report the results of \droid compared to \approach on different combination of datasets. %
Specifically, we report results for experiments similar to those carried out in Section~\ref{sec:detectiontime} as we evaluate %
its performance %
on dataset from the same year and over time. %
First, we train %
it using older dataset composed of {\tt
oldbenign} combined with one of the three
oldest malware datasets each ({\tt drebin}, {\tt 2013}, and {\tt 2014}), and test on all malware datasets. Testing on all datasets ensures the model is evaluated on dataset from the same year and newer.
With this configuration, the best result (with
{\tt 2014} and {\tt oldbenign} as training sets) is 0.62 F-measure when tested on
the same dataset. The F-measure drops to 0.33 and 0.39, respectively, when tested on samples one year into the
future and past. If we use the same configurations in \approach, 
in package mode, we obtain up to 0.97 F-measure (using {\tt 2013} and {\tt oldbenign}
as training sets), dropping to 0.73 and 0.94 respectively, one year into the future and into the past. For the datasets where \droid achieves its best result (i.e., {\tt 2014} and {\tt oldbenign}), \approach achieves an F-measure of 0.95, which drops to respectively, 0.78 and 0.93 one year into
the future and the past. The F-measure is stable even two years
into the future and the past at 0.75 and 0.92 respectively.
As a second set of experiments, we train \droid using a dataset composed of {\tt newbenign} (March 2016) combined
with one of the three most recent malware datasets each ({\tt 2014}, {\tt 2015}, and {\tt 2016}).
Again, we test \droid on all malware datasets. The best result is obtained when the {\tt 2014} and {\tt newbenign} dataset are used for both training and testing, yielding an F-measure of 0.92, which drops to 0.67 and 0.75 one year into the future and past respectively. Likewise, we use the same datasets
for \approach, with the best results achieved on the same
dataset as \droid. In package mode, \approach achieves an F-measure of 0.99 which is maintained more than
two years into the past, but drops to respectively, 0.85 and 0.81 one and two years into the future

As summarized in Table~\ref{table:droidapiresults}, \approach achieves significantly higher performance in all but one experiment than \droid.
This case occurs when the malicious training set is much older than the malicious test set.

\begin{table*}[t]
\centering
\setlength{\tabcolsep}{3pt}
\resizebox{1.005\linewidth}{!}{
\begin{tabular}{|l|cc|cc|cc|cc|cc|}
\cline{2-11}
\multicolumn{1}{c|}{} & \multicolumn{10}{c|}{\bf Testing Sets}   \\
\cline{2-11}
\multicolumn{1}{c|}{} & \multicolumn{2}{c|}{\texttt{drebin}, {\tt oldbenign}} & \multicolumn{2}{c|}{\texttt{2013}, {\tt oldbenign}} & \multicolumn{2}{c|}{\texttt{2014}, {\tt oldbenign}} & \multicolumn{2}{c|}{\texttt{2015}, {\tt oldbenign}} & \multicolumn{2}{c|}{\texttt{2016}, {\tt oldbenign}}\\
\hline %
{\bf Training Sets} &  Droid  & MaMa &  Droid   & MaMa &   Droid  & MaMa &   Droid  & MaMa &   Droid  & MaMa\\
\hline
\texttt{drebin \& oldbenign} & 0.32 & {\bf 0.96} & 0.35 & {\bf 0.95} & 0.34 & {\bf 0.72} & 0.30 & {\bf 0.39} & 0.33 & {\bf 0.42} \\
\hline
\texttt{2013 \& oldbenign} & 0.33 & {\bf 0.94} &  0.36  & {\bf 0.97} & 0.35 & {\bf 0.73} & 0.31 & {\bf 0.37} & {\bf 0.33} & 0.28 \\
\hline
\texttt{2014 \& oldbenign} & 0.36 & {\bf 0.92} & 0.39 & {\bf 0.93} & 0.62 & {\bf 0.95} & 0.33 & {\bf 0.78} & 0.37 & {\bf 0.75} \\
\hline
 & \multicolumn{2}{c|}{\texttt{drebin}, {\tt newbenign}} & \multicolumn{2}{c|}{\texttt{2013}, {\tt newbenign}} & \multicolumn{2}{c|}{\texttt{2014}, {\tt newbenign}} & \multicolumn{2}{c|}{\texttt{2015}, {\tt newbenign}} & \multicolumn{2}{c|}{\texttt{2016}, {\tt newbenign}}\\
\hline 
{\bf Training Sets} &  Droid  & MaMa &  Droid   & MaMa &   Droid  & MaMa &   Droid  & MaMa &   Droid  & MaMa\\
\hline
\texttt{2014 \& newbenign} & 0.76 & {\bf 0.98} & 0.75 & {\bf 0.98} & 0.92 & {\bf 0.99} & 0.67 & {\bf 0.85} & 0.65 & {\bf 0.81} \\
\hline
\texttt{2015 \& newbenign} & 0.68 & {\bf 0.97} & 0.68 & {\bf 0.97} & 0.69 & {\bf 0.99} & 0.77 & {\bf 0.95} & 0.65 & {\bf 0.88} \\
\hline
\texttt{2016 \& newbenign} & 0.33 & {\bf 0.96} & 0.35 & {\bf 0.98} & 0.36 & {\bf 0.98} & 0.34 & {\bf 0.92} & 0.36 & {\bf 0.92} \\
\hline
\end{tabular}
}
\caption{Classification performance of \droid (Droid)~\protect\cite{Aafer2013DroidAPIMiner} vs \approach (MaMa) in package mode using Random Forest.}
\label{table:droidapiresults}
\vspace{-0.25cm}
\vspace{0.25cm}
\end{table*}

\section{Finer-grained Abstraction}
\label{sec:classeval}
In Section~\ref{sec:markoveval}, %
we have showed that building models from abstracted API calls 
allows \approach to obtain high accuracy, %
as well as to retain it over the years, which is crucial due to the continuous evolution of the Android ecosystem.
Our experiments have focused on operating \approach in family and package mode (i.e., abstracting calls to family or package).

In this section, we investigate  whether a finer-grained abstraction -- namely, to classes -- performs better %
in terms of detection accuracy. Recall that %
our system performs better in package mode than in family mode due to the system using in the former, finer and more features to distinguish between malware and benign samples, so we set to verify whether 
one can trade-off higher computational and memory complexities for better accuracy. %
To this end, %
as discussed in Section~\ref{sec:abstraction}, we abstract each API call to its corresponding class name using a whitelist of all classes in the Android API, which consists of 4,855 classes (as of API level 24), and in the Google API, with 1,116 classes, plus self-defined and obfuscated.

\subsection{Reducing the size of the problem}
Since there are 5,973 classes, 
processing the Markov chain transitions that results in this mode increases the memory requirements. Therefore, to reduce the complexity, %
we cluster classes based on their similarity. 
To this end, we build a co-occurrence matrix that counts the number of times a class is used with other classes in the same sequence in all datasets. 
More specifically, we build a co-occurrence matrix $C$, of size (5,973$\cdot$5,973)/2, where $C_{i,j}$ denotes the number of times the i-th and the j-th  class appear in the same sequence, for all apps in all datasets. 
From the co-occurrence matrix, we compute the cosine similarity (i.e., $cos(\pmb x, \pmb y) = \frac {\pmb x \cdot \pmb y}{||\pmb x|| \cdot ||\pmb y||}$), and use k-means to cluster the classes based on their similarity into 400 clusters and use each cluster as the label for all the classes it contains. %
Since we do not cluster classes abstracted to self-defined and obfuscated, %
we have a total of 402 labels.%

\begin{table*}[t]
\centering
\footnotesize
\begin{tabular}{|l|p{0.75cm}p{0.75cm}r|p{0.75cm}p{0.75cm}r|}
\hline
\multirow{ 2}{*}{\backslashbox{\bf Dataset}{\bf \hspace{-0.2cm}Mode\hspace{-0.2cm}}}  & \multicolumn{6}{c|}{[Precision, Recall, $\textbf{F}$-measure]}   \\
\cline{2-7}
 & \multicolumn{3}{c|}{Class} & \multicolumn{3}{c|}{Package} \\
\hline
{\texttt{drebin}, {\tt oldbenign}} & 0.95 & 0.97 & 0.96	& 0.95 & 0.97 & 0.96 \\
\hline
{\texttt{2013}, {\tt oldbenign}} & 0.98 & 0.95 & 0.97 & 0.98 & 0.95 & 0.97 \\
\hline
{\texttt{2014}, {\tt oldbenign}} & 0.93 & 0.97 & 0.95 & 0.93 & 0.97 & 0.95 \\
\hline
{\texttt{2014}, {\tt newbenign}} & 0.98 & 1.00 & 0.99 & 0.98 & 1.00 & 0.99 \\ 
\hline
{\texttt{2015}, {\tt newbenign}} & 0.93 & 0.98 & 0.95 & 0.93 & 0.98 & 0.95 \\
\hline
{\texttt{2016}, {\tt newbenign}} & 0.91 & 0.92 & 0.92 & 0.92 & 0.92 & 0.92 \\
\hline
\end{tabular}
\footnotesize
\caption{\approach's Precision, Recall, and F-measure when trained and tested on dataset from the {\em same} year %
in class and package modes.}
\label{table:clusterresults}
\vspace{-1cm}
\end{table*}

\subsection{Class Mode Accuracy}
In Table~\ref{table:clusterresults}, we report the resulting F-measure in class mode using the above clustering approach when the classifier is trained and tested on samples from the same year. Once again, we also report the corresponding results from package mode for comparison (cf Section~\ref{sec:mama}). %
Overall, we find that class abstraction does not provide significantly higher accuracy.
In fact, compared to package mode, abstraction to classes only yields an average increase in F-measure of 0.0012. 

\begin{figure}[t]
\centering
\subfigure[\label{fig:compPastTest}{Newer/Older}]
{\includegraphics[width=0.38\textwidth]{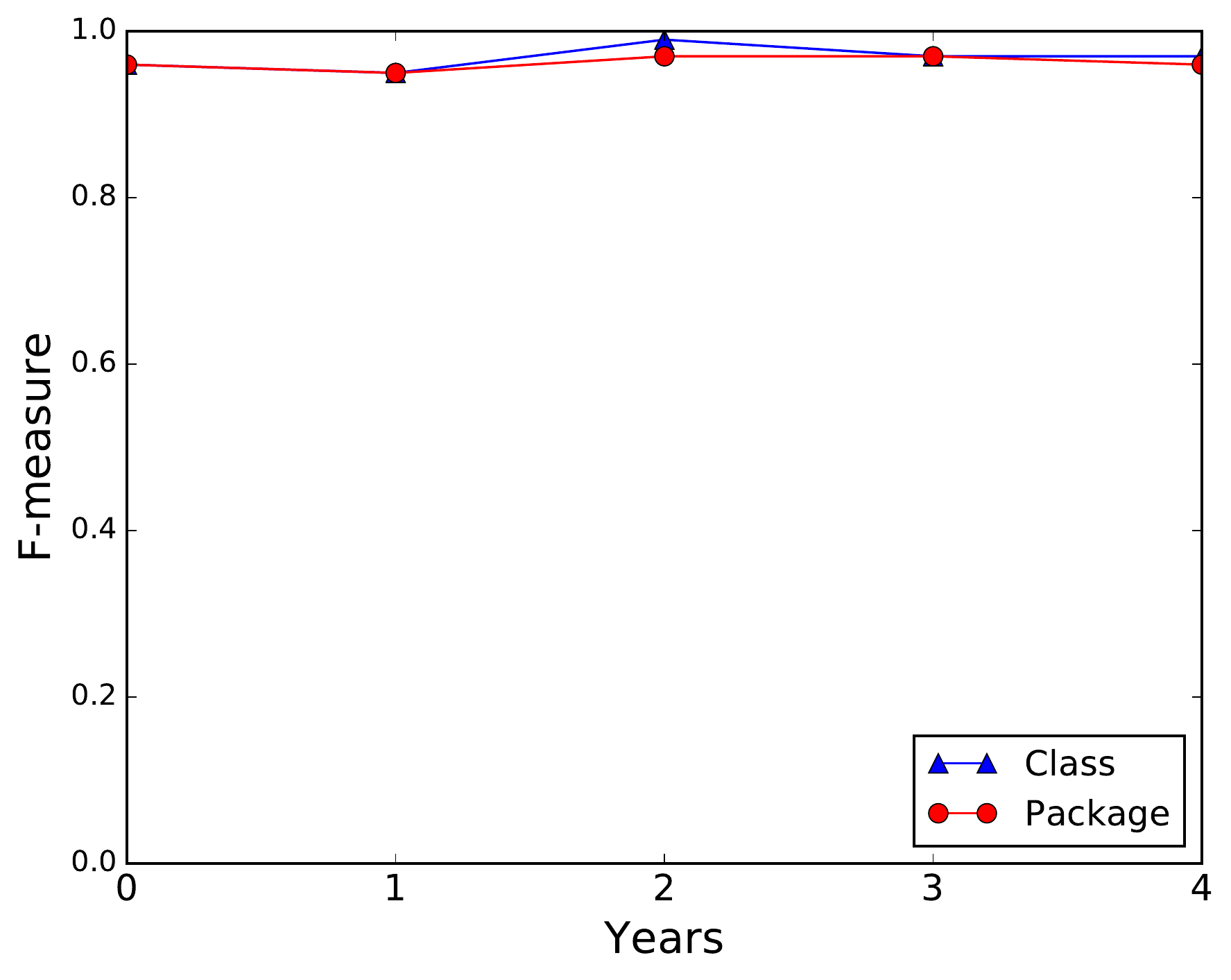}}
\hspace{0.5cm}
\subfigure[\label{fig:compFutureTest}{Older/Newer}]
{\includegraphics[width=0.38\textwidth]{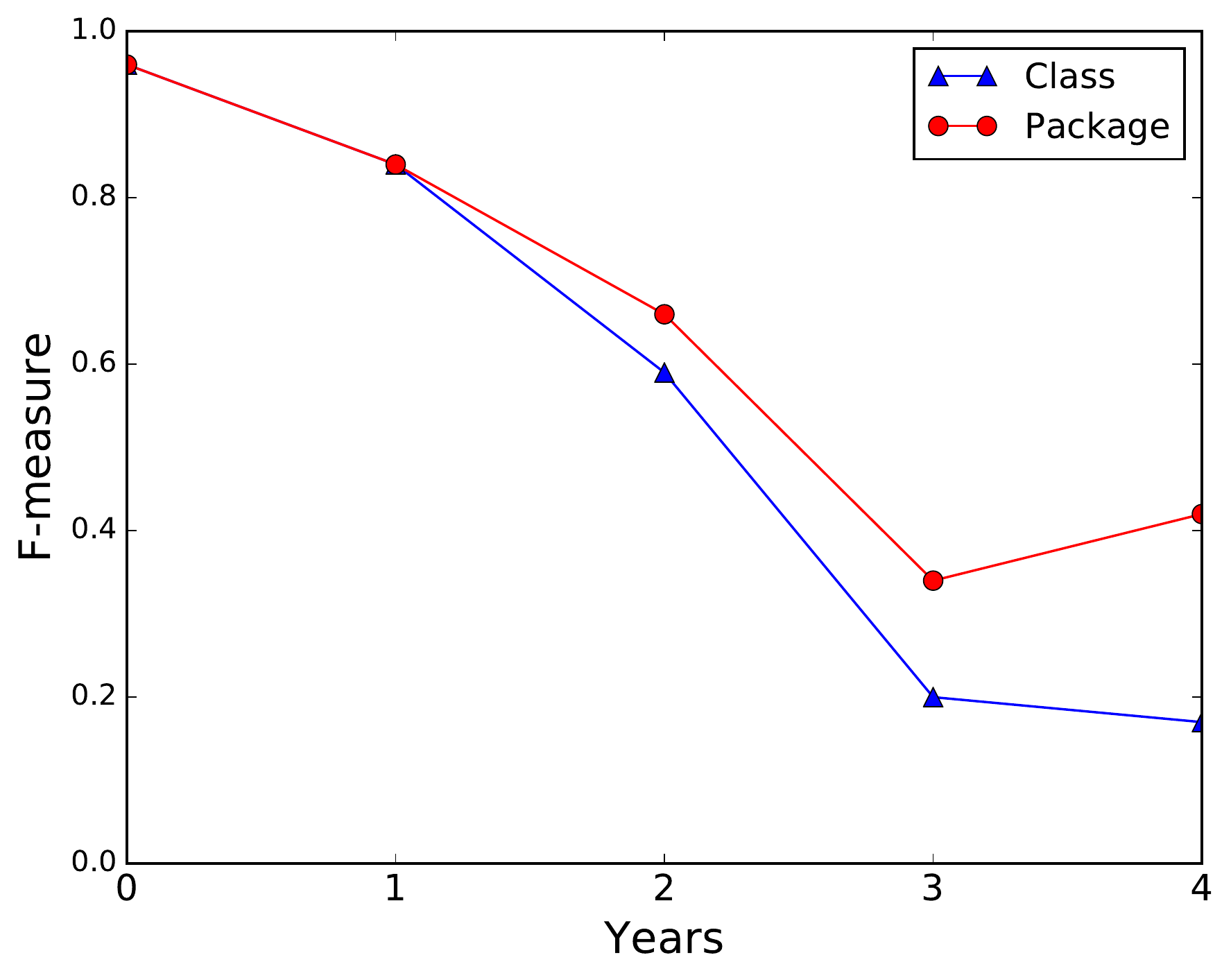}}
\vspace{-0.4cm}
\caption{F-measure achieved by \approach in class mode when using {\em newer} ({\em older}) samples for training and {\em older} ({\em newer}) samples for testing.} %
\label{fig:comp}
\end{figure}

\subsection{Detection over time}
We also %
report in Fig.~\ref{fig:comp} (the x-axis shows the difference in years between the training and testing dataset.), the 
accuracy when \approach is trained and tested on dataset from different years. %
We find that, when \approach operates in class mode, it achieves an F-measure of 0.95 and 0.99, respectively, when trained with datasets one and two years newer than the test sets, as reported in Fig.~\ref{fig:compPastTest}). Likewise, when trained on datasets one and two years older than the test set, F-measure reaches 0.84 and 0.59, respectively (see Fig.~\ref{fig:compFutureTest}). 

Overall, comparing results from Fig.~\ref{fig:PmarkovF} to %
Fig.~\ref{fig:compFutureTest}, we find that finer-grained abstraction actually performs worse with time when older samples are used for training and newer for testing. %
We note that this is due to a possible number of reasons:
1) newer classes or packages in recent API releases cannot be captured in the behavioral model of older tools, whereas families are; and %
2) evolution of malware either as a result of changes in the API or patching of vulnerabilities or presence of newer vulnerabilities that allows for stealthier malicious activities. 

On the contrary, %
Fig.~\ref{fig:FmarkovP} and \ref{fig:compPastTest} show that finer-grained abstraction performs better when the training samples are more recent than the test samples. This is because from recent samples, we are able to capture the full behavioral model of older samples. However, our results indicate there is a threshold for the level of abstraction which when exceeded, finer-grained abstraction will not yield any significant improvement in detection accuracy. This is because API calls in older releases are subsets of subsequent releases.
For instance, when the training %
samples are two years newer, \approach achieves an F-measure of 0.99, 0.97, and 0.95 respectively, in class, package, and family modes. %
Whereas when they are three years newer, the F-measure is respectively, 0.97, 0.97, and 0.96 in class, package, and family modes.

\section{Frequency Analysis Model (\FAM)}
\label{sec:evaluation}

\approach mainly relies on (1) API call abstraction, and (2) behavioral modeling via sequence of calls. As shown, it outperforms state-of-the-art Android detection techniques, such as \droid~\cite{Aafer2013DroidAPIMiner}, that are based on the frequency of non-abstracted API calls. 
In this section, we aim to assess whether \approach's effectiveness mainly stems from the API abstraction, or from the sequence modeling. To this end, we implement and evaluate a variant that uses frequency, rather than sequences, of abstracted API calls.
More precisely, we perform frequency analysis on the API calls extracted using Androguard after removing ad libraries, as also done in \droid.
In the rest of the section, we denote this variant as \FAM (Frequency Analysis Model).

We again use the datasets %
in Table~\ref{table:dataset}
to evaluate \FAM's accuracy 
when training and testing on datasets from the same year and from different years. We also evaluate how it compares to 
standard \approach. 
Although we have also implemented \FAM in class mode, we do not discuss/report results here due to space limitation.

\subsection{\FAM Accuracy}\label{freq}

\begin{table*}[t]
\centering
\footnotesize
\begin{tabular}{|l|p{0.75cm}p{0.75cm}r|p{0.75cm}p{0.75cm}r|}
\hline
\multirow{ 2}{*}{\backslashbox{\bf Dataset}{\bf \hspace{-0.2cm}Mode\hspace{-0.2cm}}}  & \multicolumn{6}{c|}{[Precision, Recall, $\textbf{F}$-measure]}   \\
\cline{2-7}
 & \multicolumn{3}{c|}{Family} & \multicolumn{3}{c|}{Package} \\
\hline
{\texttt{drebin}, {\tt oldbenign}} & - & - & - & 0.51 & 0.57 & 0.54 \\
\hline
{\texttt{2013}, {\tt oldbenign}} & - & - & - & 0.53 & 0.57 & 0.55 \\
\hline
{\texttt{2014}, {\tt oldbenign}} & 0.71 & 0.76 & 0.73 & 0.73 & 0.73 & 0.73 \\
\hline
{\texttt{2014}, {\tt newbenign}} & 0.85 & 0.90 & 0.87 & 0.88 & 0.89 & 0.89 \\
\hline
{\texttt{2015}, {\tt newbenign}} & 0.64 & 0.70 & 0.67 & 0.68 & 0.66 & 0.67 \\
\hline
{\texttt{2016}, {\tt newbenign}} & 0.51 & 0.49 & 0.50 & 0.53 & 0.52 & 0.53 \\
\hline
\end{tabular}
\caption{Precision, Recall, and F-measure (with Random Forests) of \FAM when trained and tested on dataset from the same year in family and package modes.}
\label{table:resultsWithoutMarkov}
\vspace{-0.25cm}
\vspace{0.25cm}
\end{table*}

\begin{figure}[t]
\centering
\subfigure[\label{fig:FM10fam}{family mode}]
{\includegraphics[width=0.45\textwidth]{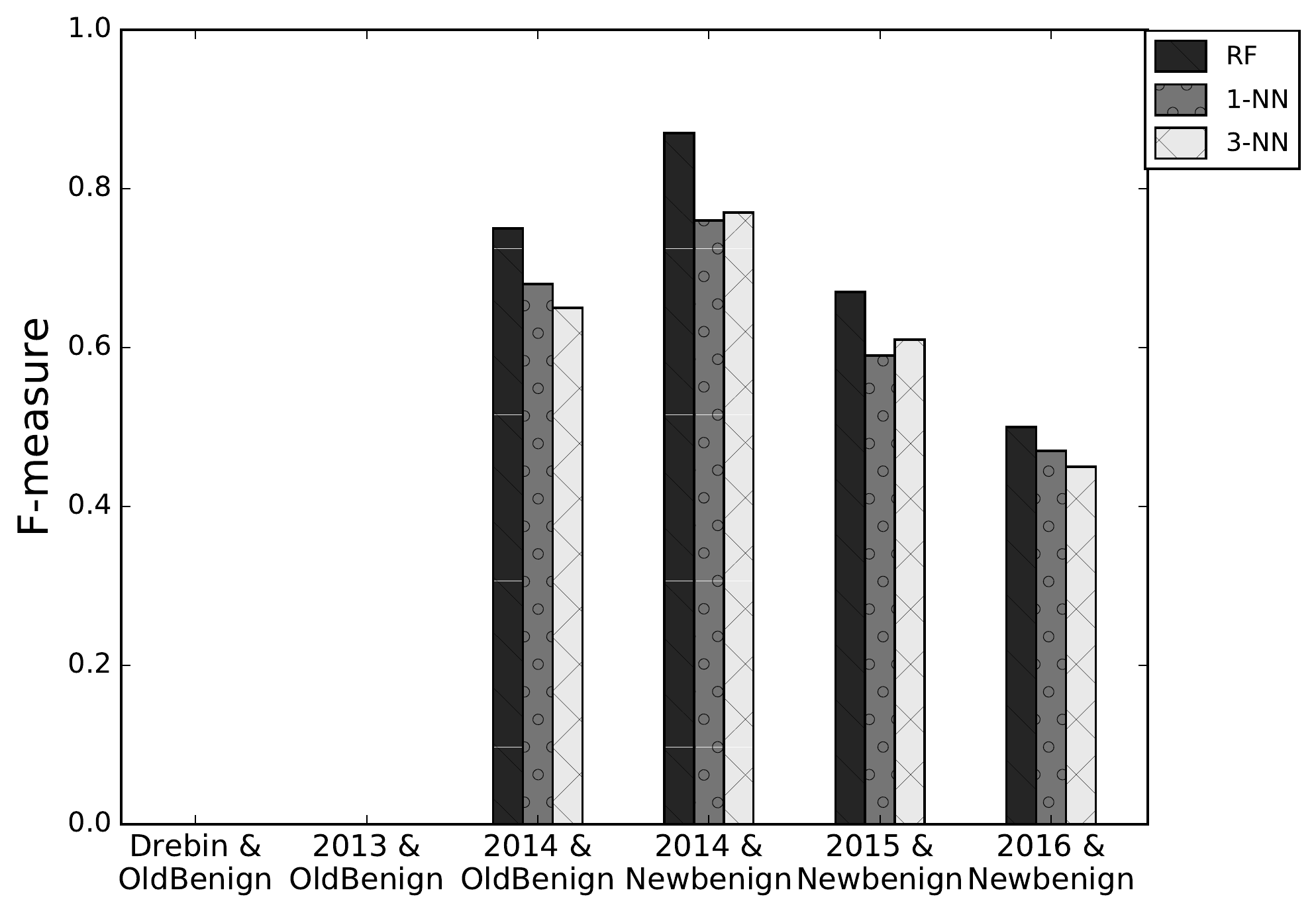}}
\subfigure[\label{fig:FM10pac}{package mode}]
{\includegraphics[width=0.45\textwidth]{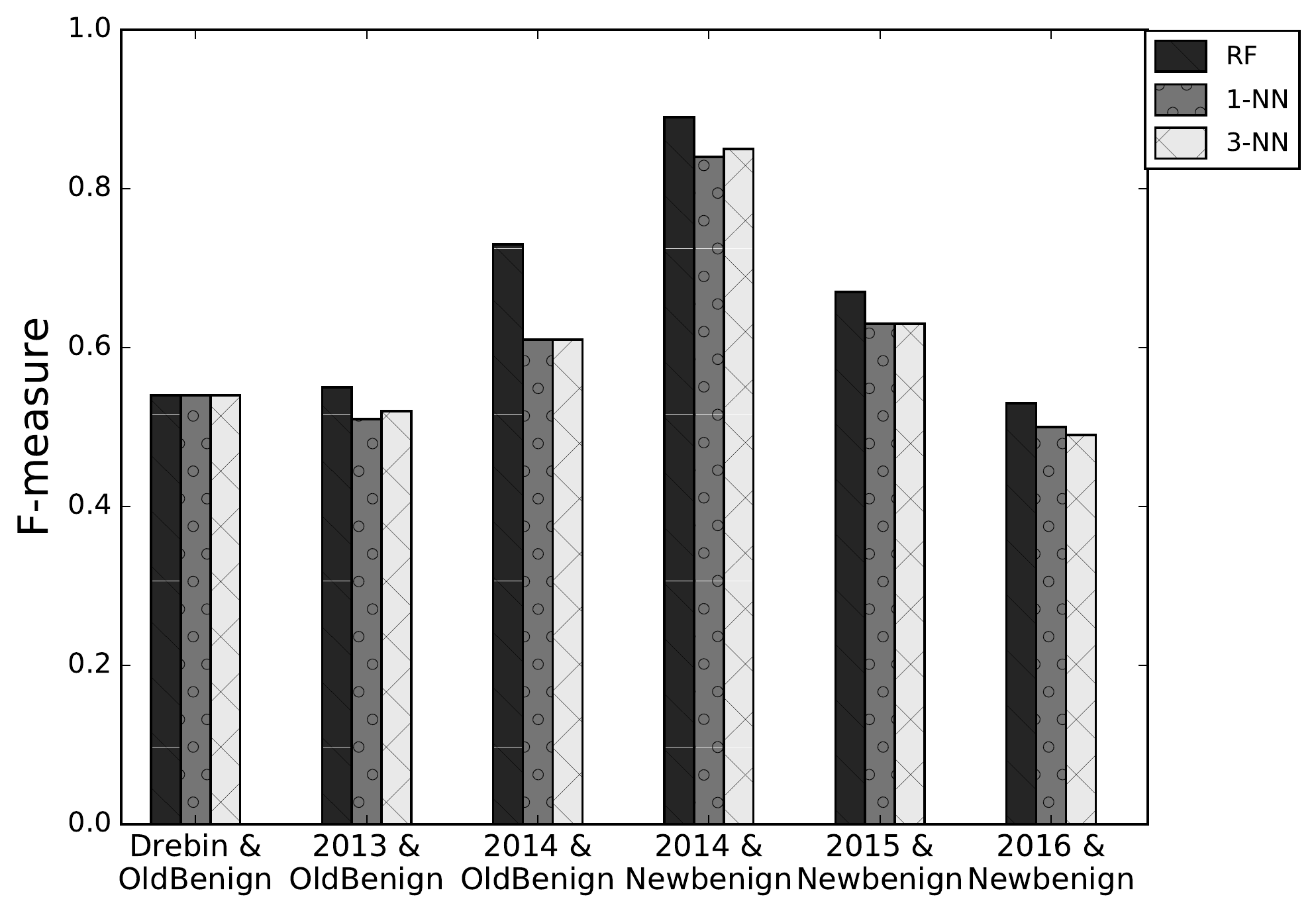}}
\vspace{-0.3cm}
\caption{F-measure for the \FAM variant, over same-year datasets, with different classifiers.}
\label{fig:FM10}
\end{figure}

We start our evaluation by measuring how well \FAM detects malware by training and testing using samples that are developed around the same time.
Fig.~\ref{fig:FM10} reports the F-measure achieved in family and package modes using three different classifiers. %
Also, Table~\ref{table:resultsWithoutMarkov} reports Precision, Recall, and F-measure achieved by \FAM on each dataset combination, when operating in family and package mode, using Random Forests. We only report the results from the Random Forest classifier because it outperforms both the 1-NN and 3-NN classifiers. %

\descr{Family mode.} Due to the number of possible families (i.e., 11), \FAM builds a model from all families that occur more in our malware dataset than the benign dataset. Note that in this modeling approach, we also remove the {\tt junit} family as it is mainly used for testing. 
When the {\tt drebin} and {\tt 2013} malware datasets are used in combination with the {\tt oldbenign} dataset, there are no families that are more frequently used in these datasets than the benign dataset. As a result, \FAM does not yield any result with these datasets as it operates by building a model only from API calls that are more frequently used in malware than benign samples. With the other datasets, there are two ({\tt 2016}), four ({\tt 2014}), and five families ({\tt 2015}) that occur more frequently in the malware dataset than the benign one.

From Fig.~\ref{fig:FM10fam}, we observe that F-measure is always at least 0.5 with Random Forests, and when tested on the {\tt 2014} (malware) dataset, it reaches 0.87. 
In general, lower F-measures are due to increased false positives.
This follows %
a similar trend observed in Section~\ref{sec:detectiontime}.

\descr{Package mode.} When \FAM operates in package mode, it builds a model using the minimum of, all API calls that occur more frequently in malware or the top 172 API calls used more frequently in malware than benign apps. We use the top 172 API calls as we attempt to build the model where possible, with packages from at least two families (the {\tt android} family has 171 packages). In our dataset, there are at least two ({\tt 2013}) and at most 39 ({\tt 2016}) packages that are used more frequently in malware than in benign samples. Hence, all packages that occur more in malware than benign apps are always used to build the model.

Classification performance improves in package mode,
with F-measure ranging from 0.53 with {\tt 2016} and {\tt newbenign} to 0.89 with {\tt 2014} and {\tt newbenign}, using Random Forests. Fig.~\ref{fig:FM10pac} shows that Random Forests generally provides better results also in this case. Similar to family mode, the {\tt drebin} and {\tt 2013} datasets respectively, have only five and two packages that occur more than in the {\tt oldbenign} dataset. Hence, the results when these datasets are evaluated %
is poor due to the limited amount of features.

\descr{Take Aways.} Although we discuss in more detail the performance of the \FAM variant vs the standard \approach in Section~\ref{sec:famvsmama}, we can already observe that the former does not yield a robust model, mostly due to the fact that in some cases, no abstracted API calls occur more in malware than benign samples. %

\subsection{Detection Over Time}

\begin{figure}[t]
\centering
\subfigure[\label{fig:famTestF}{family mode}]
{\includegraphics[width=0.38\textwidth]{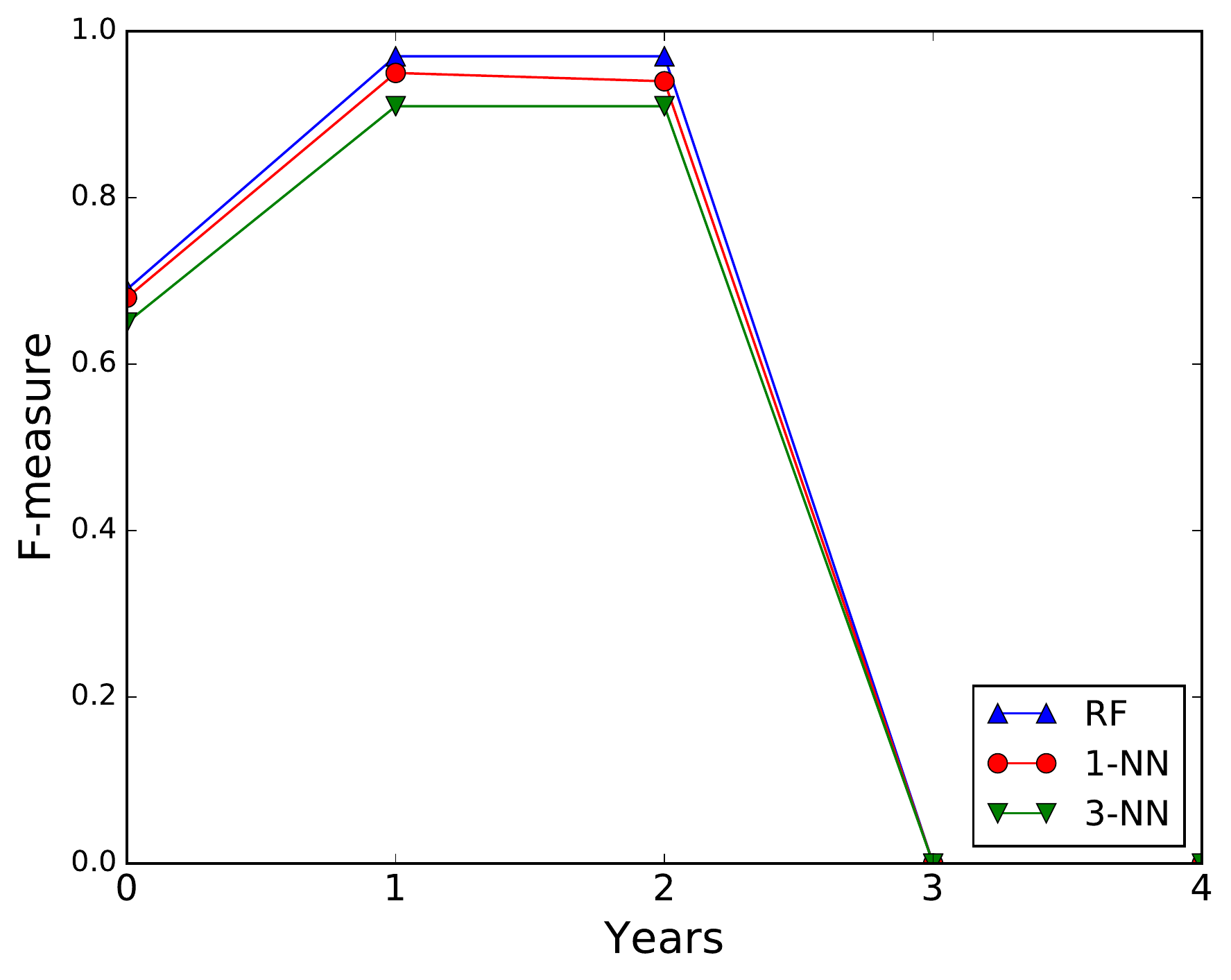}}
\hspace{0.5cm}
\subfigure[\label{fig:packTestF}{package mode}]
{\includegraphics[width=0.38\textwidth]{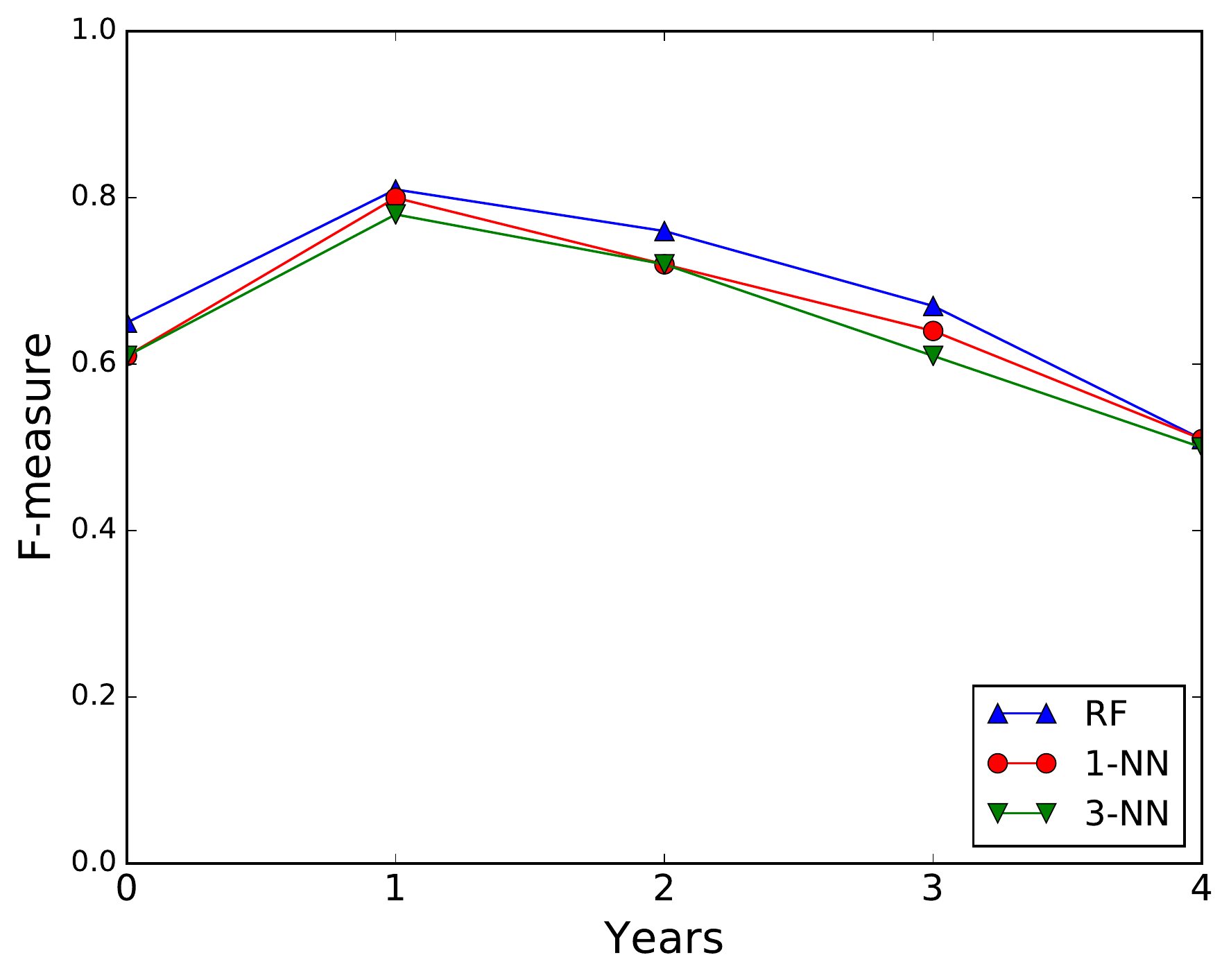}}
\vspace{-0.3cm}
\caption{F-measure achieved by \FAM using {\em older} samples for training and {\em newer} samples for testing.} %
\label{fig:FPTestF}
\end{figure}

Once again, we evaluate the detection accuracy over time, i.e., we train \FAM using older samples and test it with newer samples and vice versa. We report the F-measure as the average of the F-measure over multiple dataset combinations; e.g., when training with newer samples and testing on older samples, the F-measure after three years is the average of the F-measure when training with ({\tt 2015, newbenign}) and ({\tt 2016, newbenign}), respectively, and testing on {\tt drebin} and {\tt 2013}. %

\descr{Older training, newer testing.} In Fig.~\ref{fig:famTestF}, we show the F-measure 
when \FAM operates in family mode and is trained with datasets that are older than the classified datasets. The x-axis reports the difference in years between the training and testing
dataset. We obtain an F-measure of 0.97 when training with samples that are one year older than the samples in the testing set.
As mentioned in Section~\ref{freq}, there is no result when the {\tt drebin} and {\tt 2013} datasets are used for training, hence, after 3 years the F-measure is 0. In package mode, the F-measure %
is 0.81 after one year, and 0.76 after two (Fig.~\ref{fig:packTestF}). 

While \FAM appears to perform better in family mode than in package mode, %
note that the detection accuracy after one and two years in family mode does not include results when the training set is ({\tt drebin, oldbenign}) or ({\tt 2013, oldbenign}) %
(cf~Section~\ref{freq}). We believe this is as a result of \FAM performing best when trained on the {\tt 2014} dataset in both modes and performing poorly in package mode when trained with ({\tt drebin, oldbenign}) and ({\tt 2013, oldbenign}) due to limited features. For example, result after two years is the average of the F-measure when training with ({\tt 2014, oldbenign/newbenign}) datasets and testing on the {\tt 2016} dataset. Whereas in package mode, result is the average F-measure obtained from training with ({\tt drebin, oldbenign}), ({\tt 2013, oldbenign}), and ({\tt 2014, oldbenign/newbenign}) datasets and testing with respectively, {\tt 2014}, {\tt 2015}, and {\tt 2016}.  %

\begin{figure}[t]
\centering
\subfigure[\label{fig:famTestP}{family mode}]
{\includegraphics[width=0.38\textwidth]{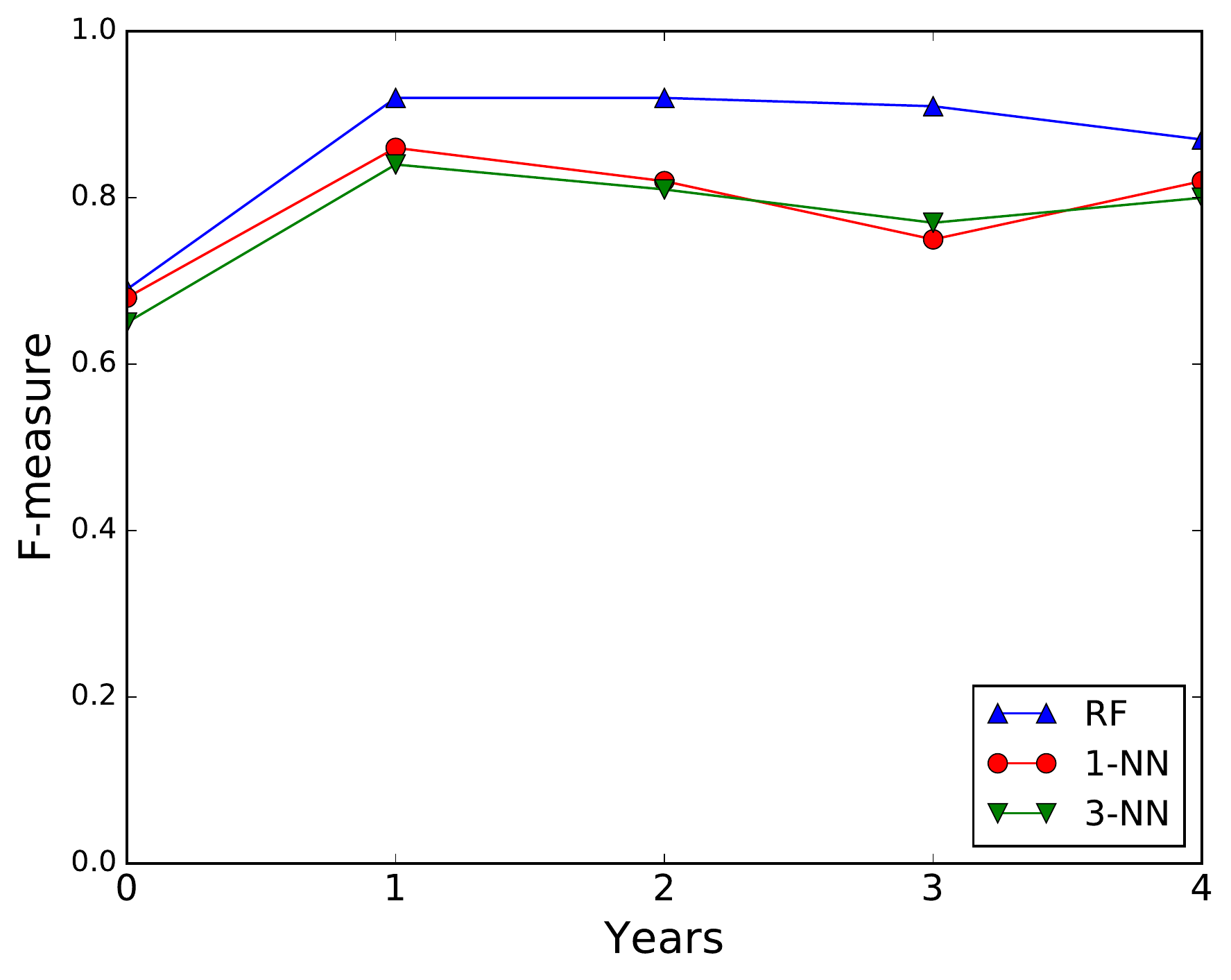}}
\hspace{0.5cm}
\subfigure[\label{fig:packTestP}{package mode}]
{\includegraphics[width=0.38\textwidth]{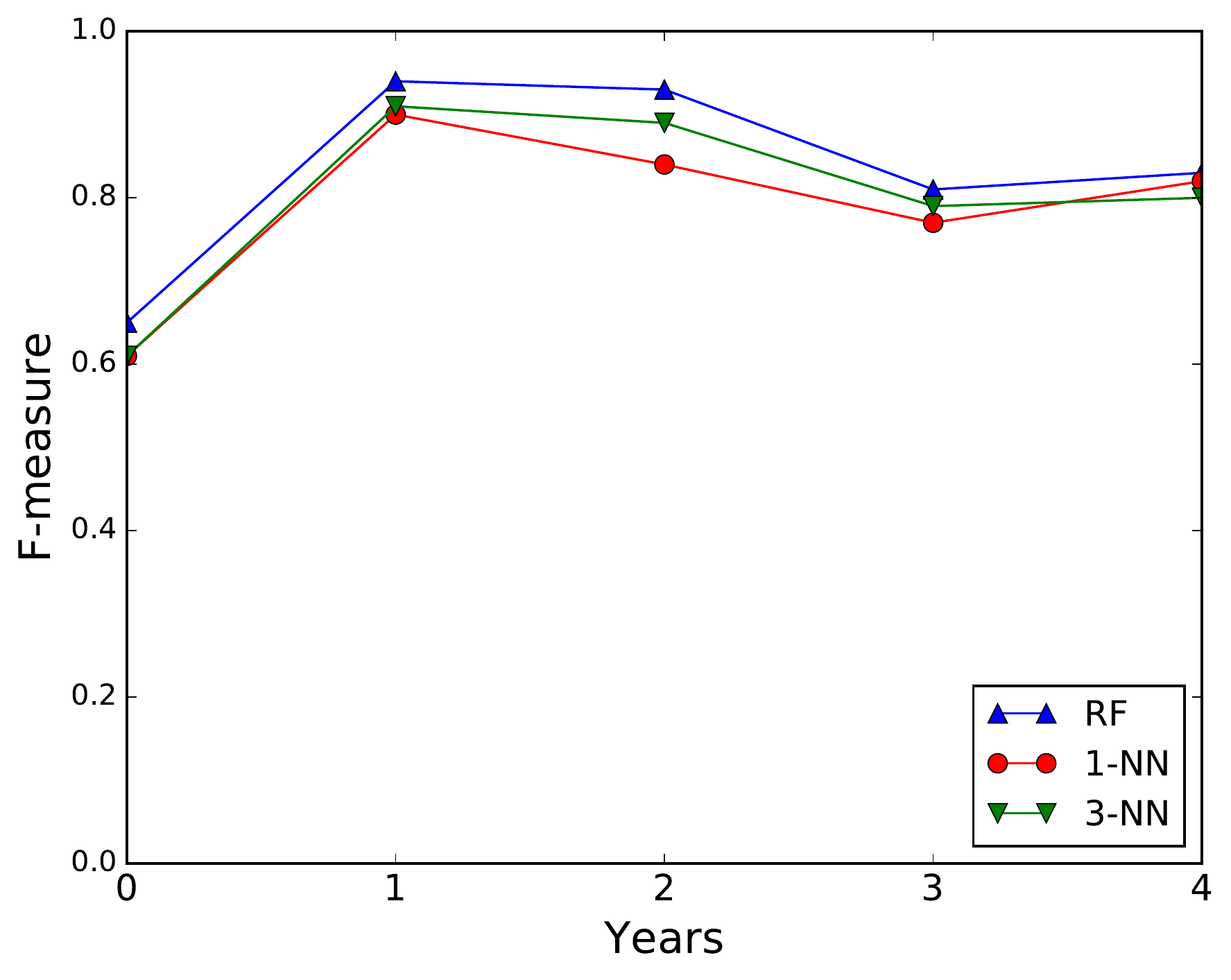}}
\vspace{-0.3cm}
\caption{F-measure achieved by \FAM using {\em newer} samples for training and {\em older} samples for testing.} %
\label{fig:FPTestP}
\end{figure}

\descr{Newer training, older testing.} We also evaluate the opposite setting, i.e., training \FAM
with newer datasets, and checking whether it is able to detect malware developed years
before. Specifically, Fig.~\ref{fig:FPTestP} %
reports results when training
\FAM with samples from a given year, and testing it with others that are up to 4 years older showing that F-measure ranges from 0.69 to 0.92 in family mode and 0.65 to 0.94 in package mode.
Recall that in family mode, \FAM is unable to build a model when {\tt drebin} and {\tt 2013} are used for training, thus, effecting the overall result. This effect is minimal in this setting since the training sets are newer than the test sets, thus, the {\tt drebin} dataset is not used to evaluate any dataset while the {\tt 2013} dataset is used in only one setting, i.e., when the training set is one year newer than the testing set.

\subsection{Comparing Frequency Analysis vs. Markov Chain Model (Package and Family Mode)}\label{sec:famvsmama}
We now compare the detection accuracy of \FAM -- a variant of \approach that is based on a
frequency analysis model --
to the standard \approach, %
which is based on a Markov chain model using the sequence of abstracted API calls.

\descr{Detection Accuracy of malware from same year.} In Table~\ref{table:famVmama}, we report the accuracy of \FAM and \approach when they are trained and tested on samples from the same year using Random Forests in both family and package modes. For completeness, we also report results from \droid, showing that \approach outperforms \FAM and \droid in all tests. Both \FAM and \droid performs best when trained and tested with ({\tt 2014} and {\tt newbenign}) with F-measures of 0.89 (package mode) and 0.92, respectively. 
Overall, \approach achieves higher F-measure compared to \FAM and \droid due to both API call abstraction and Markov chain modeling of the {\em sequence} of calls, which successfully captures the behavior of the app. In addition, \approach is more robust as with some datasets, frequency analysis fails to build a model with abstracted calls when the abstracted calls occur equally or more frequently in benign samples.

\begin{table*}[t]
\centering
\footnotesize
\begin{tabular}{|l|p{0.75cm}r|p{0.75cm}r|c|}
\hline
\multirow{ 3}{*}{\backslashbox{\bf Dataset}{\bf \hspace{-0.2cm}Mode\hspace{-0.2cm}}}  & \multicolumn{5}{c|}{$\textbf{F}$-measure}   \\
\cline{2-6}
 & \multicolumn{2}{c|}{Family} & \multicolumn{2}{c|}{Package} & \\
\cline{2-6}
 & \FAM & \approach & \FAM & \approach & \droid \\
 \hline
{\texttt{drebin}, {\tt oldbenign}} & - & {\bf 0.88} & 0.54 & {\bf 0.96} & 0.32 \\
\hline
{\texttt{2013}, {\tt oldbenign}} & - & {\bf 0.92} & 0.55 & {\bf 0.97} & 0.36 \\
\hline
{\texttt{2014}, {\tt oldbenign}} & 0.73 & {\bf 0.92} & 0.73 & {\bf 0.95} & 0.62 \\
\hline
{\texttt{2014}, {\tt newbenign}} & 0.87 & {\bf 0.98} & 0.89 & {\bf 0.99} & 0.92 \\
\hline
{\texttt{2015}, {\tt newbenign}} & 0.67 & {\bf 0.91} & 0.67 & {\bf 0.95} & 0.77 \\
\hline
{\texttt{2016}, {\tt newbenign}} & 0.50 & {\bf 0.89} & 0.53 & {\bf 0.92} & 0.36 \\
\hline
\end{tabular}
\caption{F-measure of \FAM and \approach in family and package modes %
as well as, \droid~\protect\cite{Aafer2013DroidAPIMiner} 
when trained and tested on dataset from the same year.}
\label{table:famVmama}
\vspace{-0.25cm}
\vspace{0.25cm}
\end{table*}

\descr{Detection Accuracy of malware from different years.} We also compare \FAM with \approach when they are trained and tested with datasets across several years. In Table~\ref{table:droid1ndroid2}, we report the F-measures achieved by \approach and \FAM in package mode using Random Forests, and show how they compare with \droid using two different sets of experiments.
In the first set of experiments, we train \approach, \FAM, and \droid using samples comprising the {\tt oldbenign} and one of the three oldest malware datasets ({\tt drebin}, {\tt 2013}, {\tt 2014}) each, and testing on all malware datasets. \approach and \FAM both outperform \droid in all experiments in this setting, showing that abstracting the API calls improves the detection accuracy of our systems. \FAM outperforms \approach in nine out of the 15 experiments, largely, when the training set comprises the {\tt drebin}/{\tt 2013} and {\tt oldbenign} datasets. Recall that when {\tt drebin} and {\tt 2013} malware datasets are used for training \FAM in package mode, only five and two packages, respectively, are used to build the model. It is possible that these packages are the principal components (as in PCA) that distinguishes malware from benign samples. In the second set of experiments, we train \approach, \FAM, and \droid using samples comprising the {\tt newbenign} and one of the three recent malware datasets ({\tt 2014}, {\tt 2015}, {\tt 2016}) each, and testing on all malware datasets. In this setting, \approach outperforms both \FAM and \droid in all but one experiment where \FAM is only slightly better. Comparing \droid and \FAM shows that \droid only performs better than \FAM in two out of 15 experiments. In these two experiments, \FAM was trained and tested on samples from the same year and resulted in a slightly lower Precision, thus, increasing false positives.

Overall, we find that the Markov chain based model achieves higher detection accuracy in both family and package modes when \approach is trained and tested on dataset from the same year (Table~\ref{table:famVmama}) and across several years (Table~\ref{table:droid1ndroid2}).%

\begin{table*}[t]
\centering
\setlength{\tabcolsep}{3pt}
\resizebox{1.005\linewidth}{!}{
\begin{tabular}{|l|ccc|ccc|ccc|ccc|ccc|}
\cline{2-16}
\multicolumn{1}{c|}{} & \multicolumn{15}{c|}{\bf Testing Sets}   \\
\cline{2-16}
\multicolumn{1}{c|}{} & \multicolumn{3}{c|}{\texttt{drebin}, {\tt oldbenign}} & \multicolumn{3}{c|}{\texttt{2013}, {\tt oldbenign}} & \multicolumn{3}{c|}{\texttt{2014}, {\tt oldbenign}} & \multicolumn{3}{c|}{\texttt{2015}, {\tt oldbenign}} & \multicolumn{3}{c|}{\texttt{2016}, {\tt oldbenign}}\\
\hline %
{\bf Training Sets} & Droid & FAM  & MaMa & Droid & FAM & MaMa & Droid & FAM  & MaMa & Droid & FAM  & MaMa & Droid & FAM & MaMa\\
\hline
\texttt{drebin,oldbenign} & 0.32 & 0.54 & {\bf 0.96} & 0.35 & 0.50 & {\bf 0.96} & 0.34 & 0.50 & {\bf 0.79} & 0.30 & {\bf 0.50} & 0.42 & 0.33 & {\bf 0.51} & 0.43 \\
\hline
\texttt{2013,oldbenign} & 0.33 & 0.90 & {\bf 0.93} & 0.36 & 0.55  & {\bf 0.97} & 0.35 & {\bf 0.95} & 0.74 & 0.31 & {\bf 0.87} & 0.36 & 0.33 & {\bf 0.82} & 0.29 \\
\hline
\texttt{2014,oldbenign} & 0.36 & {\bf 0.95} & 0.92 & 0.39 & {\bf 0.99} & 0.93 & 0.62 & 0.73 & {\bf 0.95} & 0.33 & {\bf 0.81} & 0.79 & 0.37 & {\bf 0.82} & 0.78 \\
\hline
 & \multicolumn{3}{c|}{\texttt{drebin}, {\tt newbenign}} & \multicolumn{3}{c|}{\texttt{2013}, {\tt newbenign}} & \multicolumn{3}{c|}{\texttt{2014}, {\tt newbenign}} & \multicolumn{3}{c|}{\texttt{2015}, {\tt newbenign}} & \multicolumn{3}{c|}{\texttt{2016}, {\tt newbenign}}\\
\hline 
{\bf Training Sets} & Droid & FAM  & MaMa & Droid &  FAM   & MaMa & Droid & FAM  & MaMa & Droid & FAM  & MaMa & Droid & FAM  & MaMa\\
\hline
\texttt{2014,newbenign} & 0.76 & {\bf 0.99} & {\bf 0.99} & 0.75 & {\bf 0.99} & {\bf 0.99} & 0.92 & 0.89 & {\bf 0.99} & 0.67 & 0.86 & {\bf 0.89} & 0.65 & 0.82 & {\bf 0.83} \\
\hline
\texttt{2015,newbenign} & 0.68 & 0.92 & {\bf 0.98} & 0.68 & 0.84 & {\bf 0.98} & 0.69 & 0.95 & {\bf 0.99} & 0.77 & 0.67 & {\bf 0.95} & 0.65 & {\bf 0.91} & 0.90 \\
\hline
\texttt{2016,newbenign} & 0.33 & 0.83 & {\bf 0.97} & 0.35 & 0.69 & {\bf 0.97} & 0.36 & 0.91 & {\bf 0.99} & 0.34 & 0.86 & {\bf 0.93} & 0.36 & 0.53 & {\bf 0.92} \\
\hline
\end{tabular}
}
\caption{F-Measure of \approach (MaMa) vs our variant using frequency analysis (\FAM) vs \droid (Droid)~\protect\cite{Aafer2013DroidAPIMiner}.}
\label{table:droid1ndroid2}
\vspace{-0.25cm}
\vspace{0.25cm}
\end{table*}

\section{Runtime Performance}\label{sec:runtime}
We now analyze the runtime performance of \approach and %
the \FAM variant, 
when operating in family, package, or class mode, as well as \droid. We run our experiments on a desktop  with a 40-core 2.30GHz CPU and 128GB of RAM, but only use 1 core and allocate 16GB of RAM for evaluation.

\subsection{\approach}
We envision \approach to be integrated in offline detection systems, e.g., run by an app store. 
Recall that \approach consists of different phases, so in the following, we review the computational overhead incurred by each of them, aiming to assess the feasibility of real-world deployment.

\approach's first step involves extracting the call graph from an apk and the complexity of this task varies significantly across apps. On average, it takes 9.2s$\pm$14 (min 0.02s, max 13m)
to complete for samples in our malware sets. %
Benign apps usually yield larger call graphs, and the average time to extract them is 25.4s$\pm$63 (min 0.06s, max 18m) per app. %
Next, we measure the time needed to extract call sequences while abstracting to
families, packages or classes depending on \approach's mode of operation. In family
mode, this phase completes in about 1.3s on average (and at most 11.0s) with both benign and malicious samples. 
Abstracting to packages takes slightly longer, due to the use of 341 packages in
\approach. On average, this extraction takes 1.67s$\pm$3.1 for malicious apps and
1.73s$\pm$3.2 for benign samples. Recall that in class mode, after abstracting to classes, we cluster the classes to a smaller set of labels due to its size. Therefore, in this mode it takes on average, 5.84s$\pm$2.1 and 7.3s$\pm$4.2 respectively, to first abstract the calls from malware and benign apps to classes and 2.74s per app to build the co-occurrence matrix from which we compute the similarity between classes. Finally, clustering and abstracting each call to its corresponding class label takes 2.38s and 3.4s respectively, for malware and benign apps. In total, it takes 10.96s to abstract calls from malware apps to their corresponding class labels 
and 13.44s %
for benign apps.

\approach's third step includes Markov chain modeling and feature vector
extraction. %
With malicious samples, it takes on average 0.2s$\pm$0.3, 2.5s$\pm$3.2, and 1.49s$\pm$2.39 (and at most 2.4s, 22.1s, and 46.10s), respectively, with families, packages, and classes, whereas with benign samples, %
it takes 0.6s$\pm$0.3, 6.7s$\pm$3.8, and 2.23s$\pm$2.74 (at most 1.7s, 18.4s, and 43.98s). 
Finally, the last step is classification, and performance depends on both the machine learning algorithm employed
and the mode of operation. More specifically, running times are affected by the number of features for the app to be classified, 
and not by the initial dimension of the call graph, or by whether the app is benign or malicious. 
Regardless, in family mode, Random Forests, 1-NN, and 3-NN all take less than 0.01s. With packages, it takes, respectively, 0.65s, 1.05s, and 0.007s per app with 1-NN, 3-NN, Random Forests. Whereas it takes, respectively, 1.02s, 1.65s, and 0.05s per app with 1-NN, 3-NN, and Random Forests in class mode.

Overall, when operating in family mode, malware and benign samples take on average, 10.7s and 27.3s, respectively, to complete the entire process, from call graph extraction to classification.  %
In package mode, the average completion times for malware and benign samples are 13.37s and 33.83s, respectively. Whereas in class mode, the average completion times are, respectively, 21.7s and 41.12s for malware and benign apps. In all modes of operation, time is mostly ($>$80\%) spent on call graph extraction. %

\subsection{\FAM}
Recall that \FAM is a variant of \approach including three phases. The first one, API calls extraction, takes 0.7s$\pm$1.5 (min 0.01s, max 28.4s) per app in our malware datasets and 13.2s$\pm$22.2 (min 0.01s, max 222s) per benign app. The second phase includes API call abstraction, frequency analysis, and feature extraction. While API call abstraction is dependent on the dataset and the mode of operation, frequency analysis and feature extraction are only dependent on the mode of operation and are very fast in all modes. In particular, it takes on average, 1.32s, 1.69s$\pm$3.2, and 5.86s$\pm$2.1, respectively, to complete a malware app in family, package, and class modes. Whereas it takes on average, 1.32s$\pm$3.1, 1.75s$\pm$3.2, and 7.32s$\pm$2.1, respectively, for a benign app in family, package, and class modes. The last phase which is classification is very fast regardless of dataset, mode of operation, and classifier used. Specifically, it takes less than 0.01s to classify each app in all modes using the three different classifiers. Overall, it takes in total 2.02s, 2.39s, and 6.56s respectively, to classify a malware app in family, package, and class modes. While with benign apps, the total is 14.52s, 14.95s, and 20.52s, respectively, in family, package, and class modes.

\subsection{\droid}
Finally, we evaluate the runtime performance of \droid~\cite{Aafer2013DroidAPIMiner}. Its first step, i.e., extracting API calls, takes 0.7s$\pm$1.5 (min 0.01s, max 28.4s) per app in our malware datasets. %
Whereas it takes on average, 13.2s$\pm$22.2 (min 0.01s, max 222s) per benign app. %
In the second phase, i.e., frequency and data flow analysis, it takes, on average, 4.2s per app. Finally, classification using 3-NN is very fast: 0.002s on average. Therefore, in total, \droid takes respectively, 17.4s and 4.9s for a complete execution on one app from our benign and malware datasets, %
which while faster than \approach, %
achieves significantly lower accuracy.
In comparison to \approach, \droid takes 5.8s and 9.9s less on average to analyze and classify a malicious and benign app when \approach operates in family mode and 8.47s and 16.43s less on average in package mode.  

\subsection{Take Aways}
In conclusion, our experiments show that our prototype implementation of \approach is scalable enough to be deployed. Assuming that, everyday, a number of apps in the order of 10,000 are submitted to Google Play, and using the average execution time of benign samples in family (27.3s), package (33.83s), and class (41.12s) modes, we estimate that it would take less than two hours %
to complete execution of all apps submitted daily in all modes, with just 64 cores. Note that we could not find accurate statistics reporting the number of apps submitted everyday, but only the total number of apps on Google Play.\footnote{\url{http://www.appbrain.com/stats/number-of-android-apps}} On average, this number increases by a couple of thousands per day, and although we do not know how many apps are removed, we believe 10,000 apps submitted every day is likely an upper bound.

\section{Discussion}
\label{sec:discussion}
We now %
discuss the implications of our results with
respect to the feasibility of modeling app behavior using static analysis and Markov chains, 
discuss possible evasion techniques, 
and highlight some limitations of our approach.

\subsection{Lessons Learned}
Our work yields important insights around the use of API calls in malicious apps, showing that, by abstracting the API calls to higher levels and modeling these abstracted calls, %
we can obtain high detection accuracy %
and retain it over %
several years,
which is crucial due to the continuous evolution of the Android ecosystem.

As discussed in Section~\ref{sec:data}, the use of API calls changes over time, and in different ways across malicious and benign samples. 
From our newer datasets, which include samples up to Spring 2016 (API level 23), we observe that
newer APIs introduce more packages, classes, and methods, while also deprecating some. %
Fig.~\ref{fig:NumCalls} %
show that benign apps use more
calls than malicious ones developed around the same time. We also notice an interesting trend in
the use of Android (Fig.~\ref{fig:FamAndro}) and Google (Fig.~\ref{fig:FamGoogle}) APIs: malicious apps follow the same trend as
benign apps in the way they adopt certain APIs, but with a delay of some years. %
This might be a side effect of Android malware authors' tendency
to repackage benign apps, adding their malicious functionalities onto them.

Given the frequent changes in the Android framework and the continuous evolution of malware, systems like \droid~\cite{Aafer2013DroidAPIMiner} -- being dependent on the presence or the use
of certain API calls -- become increasingly less effective with time. %
As shown in Table~\ref{table:droidapiresults}, %
malware that uses API calls released after those used by samples in the training set cannot be identified by these systems.
On the contrary, as shown in Fig.~\ref{fig:PmarkovF}, \approach detects malware samples that are {\em 1 year} newer than the training set obtaining an F-measure of 0.86 (as opposed to 0.46 with \droid) when the apps are modeled as Markov chains. After 2 years, the value is still at 0.75 (0.42 with \droid), dropping to 0.51 after 4 years. 

We argue that the effectiveness of \approach's classification remains relatively high ``over the years'' owing to the
Markov models capturing app's behavior. 
These models tend to be more robust to malware evolution
because abstracting to, %
e.g., packages makes the system less susceptible to the introduction of new API calls.
To verify this, we developed a variant of \approach named \FAM that abstracts API calls and is based on frequency analysis similar to \droid. Although, the addition of API call abstraction results in an improvement of the detection accuracy of the system (an F-measure of 0.81 and 0.76 after one and two years respectively), it also resulted in scenarios where there are no API calls that are more frequently used in malware than benign apps.  %

In general, abstraction allows \approach capture newer
classes/methods added to the API, since these are abstracted to already-known families or packages. As a result, it does not require any change or modification in its operation with new API releases.
In case new packages are added to new API level releases, %
\approach only requires adding a new state for each new package to the Markov chains, and the probability of a transition from a state to this new state in old apps (i.e., apps without the new packages) would be 0. %
That is, if only two packages are added in the new API release, only two states need to be added which requires trivial effort. In reality though, methods and classes are more frequently added than packages with new API releases. Hence, we also evaluate whether \approach still performs as well as in package mode when we abstract API calls to classes and measure the overall overhead increase. Results from Figure \ref{fig:comp}, \ref{fig:FPTestF}, and \ref{fig:FPTestP} indicate that finer-grained abstraction is less effective as time passes when older samples are used for training and newer samples for testing, while they are more effective when samples from the same year or newer than the test sets are used for training. 
However, while all three modes of abstraction performs relatively well, we believe abstraction to packages is the most effective as it generally performs better than family -- though less lightweight -- and as well as class but more efficient.

\subsection{Potential Machine Learning Bias} 
Recently, Pendelebury et al.~\cite{pendlebury2018tesseract} present Tesseract, which attempts to eliminate spatial and temporal bias present in malware classifiers. 
To this end, the authors suggest malware classifiers should enforce three constraints: 1) {\it temporal training consistency}, i.e., all objects in the training set must temporally precede all objects in the testing set, 2) {\it temporal goodware/malware windows consistency}, i.e., in every testing
slot of size $\Delta$, all test objects must be from the same time window, and 3) {\it realistic malware-to-goodware percentage in testing}, i.e., the testing distribution must reflect the real-world percentage of malware observed in the wild.

With respect to temporal bias, recall that we evaluate \approach over several experimental settings and many of these settings do not violate these constraints. For example, there is temporal goodware/malware windows consistency in many of the settings when it is evaluated using samples from the same year (e.g., newbenign and 2016) and when it is trained on older (resp., newer) samples and tested on newer (resp., older) samples, e.g., oldbenign and 2013. While temporal training consistency shows the performance of the classifier in detecting unknown samples (especially when the unknown samples are not derivatives of previously known malware family), many malware classifiers naturally have false negatives. Malware families in these false negatives could form the base for present or ``future'' malware. For example, samples previously classed as goodware from Google Play Store are from time to time detected as malware~\cite{bi, symantecMalware, fortuneMalware, checkpoint}.  
As a result of samples previously detected as goodware now being detected as malware, we argue that {\em robust} malware classifiers should be able to detect previous (training with newer samples and testing on older samples), present (training and testing on samples from the same time window), and future (training on older samples and testing on newer samples) malware objects effectively.

Note that we do not enforce the constraint eliminating spatial bias as proposed in Tesseract. When evaluating \approach, the minimum and maximum percentages of malware in the testing set are, resp., 49.7\% and 84.85\%, which may in theory effect Precision and Recall. However, as highlighted in~\cite{pendlebury2018tesseract}, estimating the percentage of malicious Android apps in the wild with respect to goodware, is a non-trivial task, with different sources reporting different results~\cite{lindorfer2014andradar, googleReport}. 
For more considerations on how the Tesseract framework uses \approach, please refer to~\cite{blog}.

\subsection{Evasion}
\label{sec:evasion}
Next, we discuss possible evasion techniques and how they can be addressed.
One straightforward evasion approach could be to repackage a benign app with small snippets of malicious code added to a few classes. However, it is difficult to embed malicious code in such a way that, at the same time, the resulting Markov
chain looks similar to a benign one. For instance, our running example from Section~\ref{sec:method} 
(malware posing as a memory booster app and executing unwanted commands as root) 
is correctly classified by \approach; although most functionalities in
this malware are the same as the original app, injected API calls generate some transitions in the Markov chain that are not typical of benign samples. %

The opposite procedure, i.e., embedding portions of benign code into a malicious app, is also likely ineffective against \approach, since, for each app, we derive the feature vector from the transition probability between calls over the entire app. A malware developer would have to embed benign code inside the malware in such a way that the overall sequence of calls yields similar transition probabilities as those in a benign app, but this is difficult to achieve because if the sequences of calls have to be different (otherwise there would be no attack), then the models will also be different. 

While \approach is able to detect our running example that employs this piggybacking/repacking evasion technique discussed above, it may still be possible to evade \approach using the technique. For example, as discussed in Section~\ref{sec:data}, malware samples show the same characteristics in terms of level of complexity and fraction of API calls from certain API call families as benign apps with a few years of delay. An adversary could inject more API calls into a malicious sample so as to mimic the characteristics of a benign sample, which would in turn effect the transition probabilities of the app's Markov chains and may result in misclassification. As part of future work, we plan to investigate how this evasion technique affects the effectiveness
of \approach and other API call-based malware detection tools. For example, if a benign app is repackaged and only a single method that performs malicious activity is injected, does
\approach detect the app as malicious? If more methods and classes that perform
malicious activities are added, is there a threshold of the number of methods or classes at which \approach detects an app as malicious and benign otherwise?

Attackers could also try to use reflection, dynamic code loading, or native code~\cite{poeplau2014execute} to evade \approach. Because \approach uses static analysis, 
it fails to detect malicious code when it is loaded or determined at runtime.
However, \approach can detect reflection when a method from the reflection
package ({\tt java.lang.reflect}) is executed. Therefore, we obtain the correct
sequence of calls up to the invocation of the reflection call, which may be
sufficient to distinguish between malware and benign apps. Similarly, \approach
can detect the usage of class loaders and package contexts that can be used to
load arbitrary code, but it is not able to model the code loaded. Likewise, native code that is part of the app cannot be modeled, as it is not Java and is not processed by Soot. These limitations are not specific to \approach, but common to static analysis in general, and could be possibly mitigated using \approach alongside dynamic analysis techniques.

Another approach could be using dynamic dispatch so that a class X in package A is created to extend class Y in package B with static analysis reporting a call to root() defined in Y as X.root(), whereas at runtime, Y.root() is executed. 
This can be addressed, however, with a small increase in \approach's computational cost, by keeping track of self-defined classes that extend or implement classes in the recognized APIs, and abstract polymorphic functions of this self-defined class to the corresponding recognized package, while, at the same time, abstracting as self-defined overridden functions in the class. %

Finally, identifier mangling and other forms of obfuscation could be used; aiming to obfuscate code and hide malicious actions. However, since classes in the Android framework cannot be obfuscated by obfuscation tools, malware developers can only do so for self-defined classes. \approach labels obfuscated calls as {\tt obfuscated} so, ultimately, these would be captured in the behavioral model (and the Markov chain) for the app. %
In our samples, we observe that benign apps use significantly less obfuscation than malicious apps, indicating that obfuscating a significant number of classes is not a good evasion strategy since this would likely make the sample more easily identifiable as malicious.
Malware developers might also attempt to evade \approach by naming their 
self-defined packages in such a way that they look similar to that of the {\tt android} %
or {\tt google} APIs, e.g., %
java.lang.reflect.{\em malware}. %
However, this 
is easily prevented by first abstracting to classes %
before abstracting to any further modes as we already do.%

\subsection{Limitations}\label{sec:limits}
\approach requires a sizable amount of memory in order to perform classification, when operating in package or class mode, working on more than 100,000 features per sample. 
The quantity of features, however, can be further reduced using feature selection algorithms such as
PCA. As explained in Section \ref{sec:evaluation}, when we use 10 components from the PCA, the system performs almost as well as the one using all the features; however, using PCA comes with a
much lower memory complexity in order to run the machine learning algorithms, because the number of dimensions of the features space where the classifier operates is remarkably reduced.

Soot~\cite{ValleeRai1999Soot}, which we use to extract call graphs, fails to analyze some apks. In fact, %
we were not able to extract call graphs for a fraction (4.6\%) %
of the apps in the original datasets due to scripts either %
failing to apply the {\tt jb} phase, which is used to transform Java bytecode to the primary intermediate representation (i.e., jimple) of Soot or not able to open the apk. %
Even though this does not really affect the results of our evaluation, one could avoid it
by using %
a different/custom intermediate representation for the analysis or use different tools to extract the call graphs which we plan to do as part of future work. 

In general, static analysis methodologies for malware detection on Android could fail to capture the runtime
environment context, code that is executed more frequently, or other effects stemming from user input~\cite{arp2014drebin}.
These limitations can be addressed using dynamic analysis, or by recording function calls
on a device. %
Dynamic analysis observes the live performance of the samples, recording what activity is actually
performed at runtime. Through dynamic analysis, it is also possible to provide inputs to the app and
then analyze the reaction of the app to these inputs, going beyond static analysis limits. To this end, we plan
to integrate dynamic analysis to build the models used by \approach as part of future work.

As mentioned, the injection of malicious code into benign apps may evade \approach  when the malicious functions are a little amount of API transitions.
Future work could explore this direction to understand the sensitivity of the system to the modification of benign apps.
Finally, we have not compared \approach to other static analysis-based Android malware detection tools with publicly available code (e.g., TriFlow~\cite{mirzaei2017triflow} and AppContext~\cite{Yang2015appcontext}) that employ information flow analysis, rather, to \droid and \FAM to show, respectively, the effects of modeling the behavior of an app as Markov chains from the sequence of API calls and the effects of abstraction. 

\section{Related Work}
\label{sec:related}

Over the past few years, Android security has attracted a wealth of work by
the research community. In this section, we review (i) program analysis
techniques focusing on general security properties of Android apps, and then
(ii) systems that specifically target malware on Android.\vspace{-0.15cm}

\subsection{Program Analysis} 
Previous work on program analysis applied to Android security has either used 
static or dynamic analysis and in some cases, combined both. With %
static analysis, the program's code is decompiled in order to extract features without actually running the program, usually employing tools such as Dare~\cite{Octeau2012dare} to obtain Java bytecode. %
Whereas dynamic analysis involves real-time execution of the program, typically in an emulated or protected environment.

Static analysis techniques include work by Felt et al.~\cite{Felt2011}, who analyze API calls to
identify over-privileged apps, while Kirin~\cite{Enck2009} is a system that examines permissions requested by apps to perform a lightweight certification, using a set of security rules that indicate whether or not the security configuration bundled with the app is safe.
RiskRanker~\cite{Grace2012riskranker} aims to identify zero-day Android malware
by assessing potential security risks caused by untrusted apps. It sifts through a large number of apps from Android markets and examines them to detect certain behaviors, such as encryption and dynamic code loading, which form malicious patterns and can be used to detect stealthy malware. Other methods, such as CHEX~\cite{Lu2012chex}, use
data flow analysis to automatically vet Android apps for
vulnerabilities. %
Static analysis has also been applied in the detection of data leaks and malicious data flows from Android apps~\cite{Arzt2014flowdroid,Klieber2014,Yang2013app,kim2012scandal}. 

DroidScope~\cite{Yan2012droidscope} and TaintDroid~\cite{Enck2014taintdroid}
monitor run-time app behavior in a protected environment %
to perform dynamic taint analysis. DroidScope performs dynamic taint analysis at the machine code level, while TaintDroid  monitors how third-party apps access or manipulate users' personal data, aiming to detect sensitive data leaving the system. However, as it is unrealistic to deploy dynamic analysis techniques directly on users' devices, due to the overhead they introduce, these are typically used offline~\cite{Rastogi2013,Zhou2012hey,tam2015copperdroid}. ParanoidAndroid~\cite{Portokalidis2010} employs a virtual clone of the smartphone, %
running in parallel in the cloud and replaying activities of the device -- however, even if minimal execution traces are actually sent to the cloud, this still takes a non-negligible toll on battery life.
Recently, hybrid systems like IntelliDroid~\cite{wong2016intellidroid} have also been proposed that %
serve as input generators, producing inputs specific to dynamic analysis tools. 
Other works~\cite{Ge2011,jiang2013detecting,Xia2015,Bhoraskar2014} combining static and dynamic analysis have also been proposed.

\subsection{Android Malware Detection} 

\descr{Signature-based methods.} A number of techniques have used {\em signatures} for Android malware detection. ASTROID~\cite{Feng2017} uses {\it maximally suspicious} common subgraph (MSCS) among malware samples belonging to the same family, as signatures for malware detection. It operates by first learning an MSCS that is common to all samples belonging to the same malware family, with the MSCS then used as a signature to approximately match other samples belonging to the malware family. While approximate matching helps improve the accuracy of ASTROID when tested on unknown signatures, it would introduce more false positives due to the lower similarity threshold (score of 0.5) required to classify an app as a member of a malware family. Compared to ASTROID, \approach does not require the family label of a malware sample, i.e., it operates like the unknown signatures/zero-day  version of ASTROID with high detection efficiency.

NetworkProfiler~\cite{Dai2013} also employs a signature-based method by generating network profiles for Android apps and extracting fingerprints based on such traces, 
while Canfora et al.~\cite{Canfora2016} obtain resource-based metrics (CPU, memory, storage, network) to distinguish malware activity from benign one.
StormDroid~\cite{Chen2016storm} extracts statistical features, such as
permissions and API calls, and extend their vectors to add dynamic
behavior-based features. While its experiments show that its %
solution outperforms, in terms of accuracy, other antivirus systems, it also indicates that the quality of its detection model critically depends on the availability of representative benign and malicious apps for training~\cite{Chen2016storm}. MADAM~\cite{saracino2016madam} also extract features at four layers which are used to build a behavioral model for apps and uses two parallel classifiers to detect malware.
Similarly, ScanMe Mobile~\cite{Zhang2016} uses the Google Cloud Messaging Service (GCM) 
to perform static and dynamic analysis on apks found on the device's SD card.  %

\descr{Sequence of calls.} The sequences of system calls have also been used to detect malware in both desktop and Android environments.
Hofmeyr et al.~\cite{hofmeyr1998intrusion} show that
short sequences of system calls can be used as a signature to discriminate
between normal and abnormal behavior of common UNIX programs. Like signature-based
methods, however, these can be evaded by polymorphism and obfuscation, or
by call re-ordering attacks~\cite{kolbitsch2009effective}, even though
quantitative measures, such as similarity analysis, can be used to address some
of these attacks~\cite{Shankarapani2011}. \approach inherits the spirit of these
approaches, with a statistical method to model app behavior that is
more robust against evasion attempts.

Specific to Android, Canfora et al.
\cite{Canfora2015} %
use the sequences of three system calls (extracted from the execution traces of
apps under analysis) to detect malware. %
This approach models specific malware families, aiming to identify additional
samples belonging to such families. %
By contrast, \approach's goal is to detect previously-unseen malware, and we
also show that our system can still detect new malware %
samples that appear years after the system has been trained. %
In addition, using strict sequences of system or API calls could more easily be evaded
by malware via unnecessary calls %
to effectively evade detection. Conversely, \approach builds a behavioral model of an Android app, 
which makes it robust to this type of evasion. \approach is more scalable as it uses static analysis compared to~\cite{Canfora2015} which extracts the sequence of calls from executing each app for 60 secs on a device. As~\cite{Canfora2015} is based on system calls extracted dynamically, their work differ from \approach as the sequence of three system calls as used in the former could potentially be mapped to a single API call in the latter or insufficient for mapping to a single API call and hence, not considered to be a sequence in the latter. %

TriFlow~\cite{mirzaei2017triflow} triage risky apps by using the observed and possible information flows from sources to sinks in the apps to prioritize the apps that should be investigated further. It employs speculative rather than the actual information flows to predict the existence of information flows from sources to sinks. Compared to TriFlow, \approach models the behavior of an app via the sequence of all API calls extracted statically, rather than via flows from sources to sinks, while also been twice as fast. In addition, \approach is designed to detect malware irrespective of the malware family, whereas TriFlow may not be able to detect malware families that do not require information flow from sources to sink to act maliciously, e.g.,  ransomeware.

AppContext~\cite{Yang2015appcontext} models the context of security-sensitive behaviors (permission-protected methods, methods used for reflection and dynamic code loading, and sources and sinks methods) as a tuple of activation events or environmental attributes to differentiate between malware and benign apps. Whereas AppContext models the behavior of an app by building call graphs to specific API calls (those defined as security-sensitive), \approach takes a holistic approach, capturing all API calls. Due to this targeted characteristic of AppContext, it is about 10 times slower than \approach, while also being less effective with F-measure of 0.9 when the complete context is used compared to 0.96/0.97 achieved by \approach (with the {\tt drebin}/{\tt 2013} and {\tt oldbenign} datasets, which are around the same timespan as that used by AppContext).

\descr{Dynamic Analysis.} Dynamic analysis has also been applied to detect Android malware by using predefined scripts of common inputs that will be performed when the device is running. However, this might be inadequate due to the low probability of triggering malicious behavior, and can be side-stepped by knowledgeable adversaries, as suggested by Wong and Lie~\cite{wong2016intellidroid}. Other input approaches include random fuzzing~\cite{Machiry2013,Ye2013} and concolic testing~\cite{Anand2012,Godefroid2005}. Dynamic analysis can only detect malicious activities if the code exhibiting malicious behavior is actually running during the analysis. Moreover, according to Vidas and Christin~\cite{Vidas2014}, mobile malware authors often employ emulation or virtualization detection strategies to change malware behavior and eventually evade detection.
Also related to \approach is \textsc{AuntieDroid}~\cite{onwuzurike2018family}, which applies \approach's technique in a dynamic analysis setting by modeling the behavior of apps using traces produced from executing the apps in a virtual device.

\descr{Machine Learning.} Machine learning techniques have also been applied to assist Android malware detection. %
Chen et al.~\cite{chen2018automated} proposed {\sc Kuafudet} that uses a two-phase learning process to enhance detection of malware that attempts to sabotage the training phase of machine learning classifiers. Also, Jordaney et al.~\cite{jordaney2017transcend} proposed Transcend which is a framework for identifying aging machine learning malware classification models, i.e., using statistical metrics to compare the samples used for training a model to new unseen samples to predict degradation in the detection accuracy of the model. 
Recently, MalDozer~\cite{karbab2018maldozer} applies deep learning on the sequence of API methods, following \approach's approach. While MalDozer also discusses its effectiveness over time, it is more susceptible to changes to the Android API framework due to its use of API method calls which are sometimes deprecated with new releases.
Hou et al. introduced HinDroid~\cite{HinDroid}, %
which represents apps and API calls as a structured heterogeneous information network, and aggregates the similarities among apps
using multi-kernel learning.

Note that we have also experimented with deep learning, finding that Random Forests and k-NN perform better. 
We believe this might be due to the fact that deep learning derives its own features used in distinguishing between the classes as compared to \approach, which is designed to use statistical methods such as Markov chains.

\descr{App's Manifest.} Features such as permissions, intent filters, etc. have also been used to distinguish between malicious and benign apps. Droidmat~\cite{Wu2012Droidmat} uses API call tracing and manifest files to learn features for malware detection, Teufl et al.~\cite{teufl2016malware} apply knowledge discovery processes and lean statistical methods on app metadata extracted from the app market, while~\cite{Gascon2013} rely on embedded call graphs. DroidMiner~\cite{yang2014droidminer} studies the program logic of sensitive Android/Java framework API functions and resources, and detects malicious behavior patterns. MAST~\cite{Chakradeo2013} statically analyzes apps using features such as permissions, presence of native code, and intent filters and measures the correlation between multiple qualitative data. %

Crowdroid~\cite{Burguera2011Crowdroid} relies on crowdsourcing to distinguish between malicious and benign apps by monitoring system calls,  
while RevealDroid~\cite{garcia2015obfuscation} employs supervised learning and obfuscation-resilient methods targeting API usage and intent actions to identify their families. 
{\sc Drebin}~\cite{arp2014drebin} deduces detection patterns and
identifies malicious software directly on the device, performing a broad static
analysis. This is achieved by gathering numerous features from the manifest file
as well as the app's source code (API calls, network addresses, permissions). 
Malevolent behavior is reflected in
patterns and combinations of extracted features from the static analysis: for
instance, the existence of both SEND\_SMS %
permission and the android.hardware.telephony component in an app might indicate an attempt 
to send premium SMS messages, and this combination can eventually
constitute a detection pattern. %
\descr{\droid.} In Section~\ref{sec:compare}, we have already compared against~\cite{Aafer2013DroidAPIMiner}. This system relies on the top-169 API calls that are used more frequently in the malware than in the benign set, along with data flow analysis
on calls that are frequent in both benign and malicious Android apps, but occur up to
6\% more in the latter. As shown in our evaluation, using the most common calls observed
during training requires constant retraining, due to the evolution of both
malware and the Android API. On the contrary, \approach can effectively model
both benign and malicious Android apps, and perform an efficient classification
on them. Compared to \droid, our approach is more resilient to changes in the
Android framework, resulting in a less frequent need to re-train the classifier. %
Overall, compared to both {\sc Drebin}~\cite{arp2014drebin} and \droid~\cite{Aafer2013DroidAPIMiner}, \approach is more generic and robust as its statistical modeling does not depend on specific app characteristics, but can actually be run on any app created for any Android API level. 

\descr{Markov Chains.} Finally, Markov-chain based models for Android
malware detection like that proposed by Chen et al.~\cite{Chen2014Hmm} dynamically analyze 
system- and developer-defined actions from intent messages (used by app components to communicate with each other at runtime),
and probabilistically estimate whether an app is performing benign or malicious actions at run time, but 
obtain low accuracy overall. Canfora et al.~\cite{Canfora2016HMM} follow two approaches in the detection of malware: 1) a Hidden Markov model (HMM) to
identify %
known malware families, whereas 
\approach is designed to %
detect previously unseen malware, %
irrespective of the family; 2) a {\em structural entropy} that compares the similarity between the byte distribution of the executable file of samples belonging to the same malware families.

\section{Conclusion}
\label{sec:conclusion}

This paper presented \approach, an Android malware detection system that is based on %
modeling the sequences
of API calls as Markov chains. %
Our system is designed to operate in one of three modes, with different granularities,
by abstracting API calls to either families, packages, or classes. 
We ran an extensive experimental evaluation using, to the best of our knowledge, one of the largest malware datasets %
in an Android malware detection research paper, aiming at assessing both the accuracy of the classification 
(using F-measure, Precision, and Recall) and runtime performances.
We showed that \approach effectively detects unknown malware samples developed %
around the same time
as the samples on which it is trained (F-measure up to 0.99).
It also maintains good detection performance: one year after the model has been trained with an F-measure %
of 0.86, and 0.75 after two years. %

We compared \approach to \droid~\cite{Aafer2013DroidAPIMiner}, a
state-of-the-art system based on API calls frequently used by malware, showing that, not only does \approach outperforms \droid when trained and tested on %
datasets from the same year, but that it is also much more resilient over the years to changes in the Android API. %
We also developed a variant of \approach, called \FAM, that performs API call abstraction but is based on frequency analysis to evaluate whether \approach's high detection accuracy is based solely on the abstraction. %
We found that \FAM improves on \droid but, while abstraction is important for high detection rate and resilience to API changes, abstraction and a modeling approach based on frequency analysis is not as robust as \approach, especially in scenarios where API calls are not more frequent in malware than in benign apps.

Overall, our results demonstrate that %
statistical behavioral models introduced by \approach---in particular, abstraction and Markov chain modeling of API call sequence---are more robust than traditional techniques, highlighting how our work can form the basis of more advanced detection systems in the future.
As part of future work, we plan to further investigate the resilience to possible evasion techniques, focusing on repackaged malicious apps
as well as injection of API calls to maliciously alter Markov models. We also plan to explore %
the possibility of seeding the behavioral modeling performed by \approach with dynamic instead of static analysis. %

\descr{Acknowledgments.} We wish to thank %
Yousra Aafer for sharing the \droid source code and Yanick Fratantonio for his comments on an early draft of the paper. This research was supported by an EPSRC-funded ``Future Leaders in Engineering and Physical Sciences'' award and a small grant from GCHQ. Lucky Onwuzurike was funded by the Petroleum Technology Development Fund (PTDF), while Enrico Mariconti was supported by the EPSRC under grant 1490017.

\bibliographystyle{abbrv}
\bibliography{androidmalware}

\end{document}